\documentclass[aps,prl,showpacs,twocolumn,superscriptaddress,floatfix]{revtex4-2}
\usepackage[utf8]{inputenc}
\newcommand{\figurescale}{1}
\usepackage{verbatim}
\usepackage{amsmath}
\usepackage{mathtools}

\DeclarePairedDelimiterX\braket[2]{\langle}{\rangle}{#1 \delimsize\vert #2}

\usepackage{graphicx}
\usepackage[separate-uncertainty=true]{siunitx}
\DeclareSIUnit{\rpm}{rpm}

\usepackage[colorlinks=true,urlcolor=blue,linkcolor=blue,citecolor=blue]{hyperref}
\usepackage[usenames,dvipsnames]{xcolor}
\usepackage[T1]{fontenc}
\usepackage{placeins}
\usepackage{nicefrac}
\usepackage{braket}
\usepackage{pdfcomment}
\usepackage{xcolor}
\usepackage{braket}
\usepackage[normalem]{ulem}

\usepackage{lineno}

\usepackage{booktabs}

\usepackage{titlesec}
\titleformat{\section}{\normalfont\normalsize\bfseries\centering}{\thesection}{1em}{}
\titleformat{\subsection}{\normalfont\normalsize\bfseries\centering}{\thesubsection}{1em}{}
\titleformat{\subsubsection}{\normalfont\normalsize\bfseries\centering}{\thesubsubsection}{1em}{}

\begin{document}


\title{Growth-controlled suppression of electrically active defects in CrSBr}
%
%
\author{Sara~R.~Tulchinsky}
\affiliation{Department of Materials Science and Engineering, Massachusetts Institute of Technology, Cambridge, Massachusetts 02139, USA}
\affiliation{Department of Physics and Astronomy, Wellesley College, Wellesley, Massachusetts 02481, USA}
\author{Sergii~Grytsiuk}
\affiliation{Faculty of Physics, Bielefeld University, 33501 Bielefeld, Germany}
\affiliation{Institute for Molecules and Materials, Radboud University, Heijendaalseweg 135, 6525AJ Nijmegen, The Netherlands}
\author{Shen~van~Hassel}
\affiliation{Faculty of Physics, Bielefeld University, 33501 Bielefeld, Germany}
\affiliation{Institute for Molecules and Materials, Radboud University, Heijendaalseweg 135, 6525AJ Nijmegen, The Netherlands}
\author{Iva~Plutnarová}
\affiliation{Department of Inorganic Chemistry, University of Chemistry and Technology Prague, Technická 5, 166 28 Prague 6, Czech Republic}
\author{Rami~Dana}
\affiliation{Department of Materials Science and Engineering, Massachusetts Institute of Technology, Cambridge, Massachusetts 02139, USA}
\author{David~Sedmidubský}
\affiliation{Department of Inorganic Chemistry, University of Chemistry and Technology Prague, Technická 5, 166 28 Prague 6, Czech Republic}
\author{Zdenek~Sofer}
\affiliation{Department of Inorganic Chemistry, University of Chemistry and Technology Prague, Technická 5, 166 28 Prague 6, Czech Republic}
\author{Malte~Rösner}
\affiliation{Faculty of Physics, Bielefeld University, 33501 Bielefeld, Germany}
\affiliation{Institute for Molecules and Materials, Radboud University, Heijendaalseweg 135, 6525AJ Nijmegen, The Netherlands}
\author{Frances~M.~Ross}
\affiliation{Department of Materials Science and Engineering, Massachusetts Institute of Technology, Cambridge, Massachusetts 02139, USA}
\author{Julian~Klein}\email{jpklein@mit.edu}
\affiliation{Department of Materials Science and Engineering, Massachusetts Institute of Technology, Cambridge, Massachusetts 02139, USA}
\date{\today}
%
%
%
%
%
\begin{abstract}\textbf{
In CrSBr, as in many crystalline materials, the type and density of defects are expected to strongly influence material behavior. Identifying the underlying atomic defect configurations and controlling their populations during growth are therefore important steps toward understanding and ultimately tailoring its rich magneto-electrical properties. However, systematic control of defects in CrSBr during chemical vapor transport (CVT) growth has not yet been established. Here, we correlate CVT growth conditions with defect concentrations measured using conductive atomic force microscopy (CAFM). We focus on a characteristic defect with a strong electronic fingerprint, labeled $D^*$, and decrease its concentration by up to an order of magnitude through optimized growth conditions. We show that defect densities can be tuned by adjusting precursor stoichiometry, where sulfur- and bromine-rich conditions suppress defect formation, and by lowering the absolute growth temperatures while maintaining the same temperature gradient. Thermodynamic modeling and density functional theory calculations suggest that $D^*$ is most consistent with a sulfur-related vacancy complex rather than an isolated point defect. These results provide practical strategies for growing high-quality CrSBr with controlled defect densities.
}
\end{abstract}

%
%
\maketitle
%
%

\textbf{Introduction.} Defects are ubiquitous in materials and play a central role in determining their physical properties. Their deterministic introduction, recently demonstrated in the magnetic semiconductor CrSBr~\cite{Klein2026}, offers a route to tailoring material behavior. CrSBr~\cite{Katscher.1966,Beck.1990,Gser.1990} has attracted considerable attention due to its rich magneto-electronic phenomena, including magneto-excitons~\cite{Wilson.2021,Telford.2020}, exciton-magnon coupling~\cite{Bae.2022,Dirnberger2023}, and magneto-transport~\cite{Telford.2022,Boix-Constant.2022}. Its quasi-1D electronic structure~\cite{Klein.2023,Wu.2022} and high air stability~\cite{Torres.2023} further make it a promising material for studying few-layer and defect physics in bulk crystals~\cite{Klein.2024b}.

Charge-transport~\cite{Telford.2020,Telford.2022,Wu.2022,Chou2025}, exciton spectroscopy~\cite{TabatabaVakili2024}, and ARPES measurements~\cite{Bianchi.2023,Bianchi2023b,Watson.2024,Smolenski2025,Biktagirov2025_CrSBr} consistently show intrinsic n-type doping in CrSBr. This doping and its sample-to-sample variation complicate experimental determination of the band gap~\cite{Bianchi.2023,Bianchi2023b,Watson.2024,Smolenski2025,Biktagirov2025_CrSBr}, as band-gap renormalization due to changes in screening can be significant~\cite{Sahoo2025,Steinhoff2014}. Recent theoretical work suggests that excess Cr interstitials and Br vacancies may contribute to the n-type character~\cite{Biktagirov2025_CrSBr}. Native defects therefore limit both the determination of intrinsic material properties and the ability to isolate the effects of deliberately introduced defects~\cite{Klein2026}.

\begin{figure*}
\scalebox{\figurescale}{\includegraphics[width=1\linewidth]{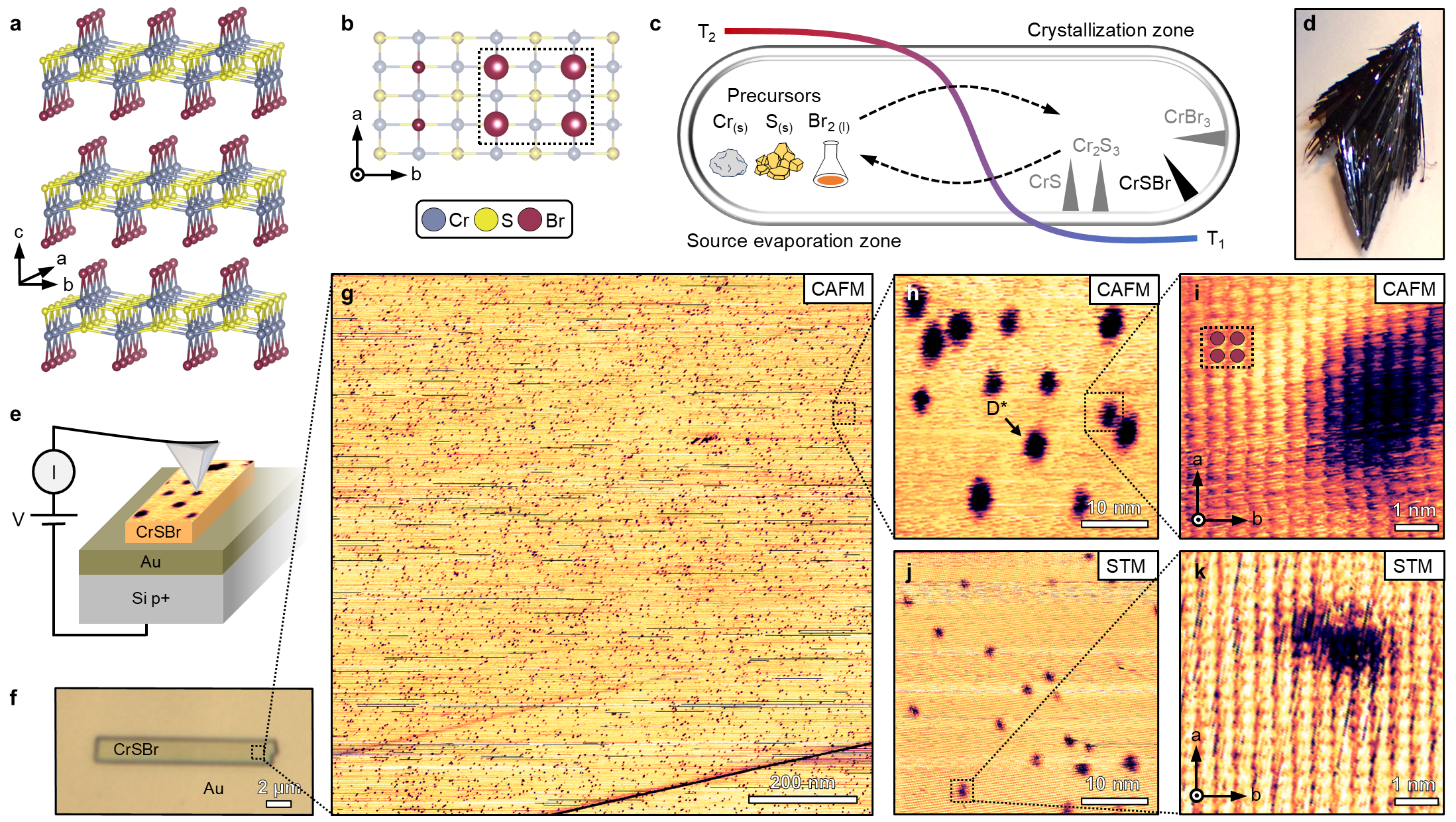}}
\caption{\textbf{Defect detection in bulk CrSBr by conductive atomic force microscopy.}
\textbf{a}, Schematic crystal structure of CrSBr.
\textbf{b}, Top view of the CrSBr lattice, with the unit cell and surface Br atoms highlighted.
\textbf{c}, Schematic of the CVT growth ampoule, showing the starting elements Cr(s), S(s), and Br$_2$(l), the applied temperature gradient ($\Delta T = T_2 - T_1$), and the formation of CrSBr together with possible secondary phases, including Cr$_2$S$_3$, CrS, and CrBr$_3$.
\textbf{d}, Optical micrograph of a bulk CrSBr crystal.
\textbf{e}, Schematic illustration of the CAFM measurement geometry and the electronic contrast associated with defects in CrSBr.
\textbf{f}, Optical micrograph of an exfoliated CrSBr flake on a Au substrate.
\textbf{g}, Large-area ($1 \times 1~\si{\micro\meter}$) CAFM current map.
\textbf{h}, Higher-resolution CAFM current map showing a characteristic defect, labeled $D^*$.
\textbf{i}, Atomic-scale CAFM current map of a single defect with the surface Br lattice overlaid.
\textbf{j}, STM topographic image showing characteristic defects, acquired with a tunneling current of $\SI{50}{\pico\ampere}$ and a sample bias of $\SI{150}{\milli\volt}$.
\textbf{k}, Atomic-scale STM topographic image of a single defect with the surface Br lattice overlaid, acquired with a tunneling current of $\SI{50}{\pico\ampere}$ and a sample bias of $\SI{150}{\milli\volt}$.}
\label{fig1}
\end{figure*}

CrSBr is grown by chemical vapor transport (CVT)~\cite{Telford.2020,Wu.2022,Klein.2022,Liu2022,Ziebel2024,Klein.2024b} and hosts a variety of defects. It has a layered FeOCl-type structure and crystallizes in the orthorhombic Pmmn space group (No. 59) (Fig.~\ref{fig1}a). Protruding Br atoms create interstitial sites in the van der Waals gap, where displaced Cr atoms can induce structural phase transformations~\cite{Klein.2022}. Atomic-scale studies have revealed point vacancies, interstitial defect complexes, and extended 1D vacancy lines~\cite{Weile2025}, while focused electron irradiation has enabled deterministic generation of user-selected interstitial defect complexes over mesoscopic scales~\cite{Klein2026}. Surface Br atoms (Fig.~\ref{fig1}b) are chemically labile under ambient conditions, particularly in the presence of water~\cite{Torres.2023}, and high Br-vacancy concentrations have been observed~\cite{Klein.2022a,Weile2025}, consistent with their low calculated formation energies~\cite{Klein.2022a,Long2023,Weile2025}.

Despite the rapidly growing interest in CrSBr, the microscopic nature, electronic role, and statistical distribution of intrinsic defects remain poorly understood. Optimized growth conditions are therefore essential for suppressing defect formation, accessing intrinsic electronic and optical properties, and providing a controlled host for deterministic defect engineering~\cite{Klein2026}.


Here, we correlate CVT growth conditions with large-area statistical measurements of defect concentrations using conductive atomic force microscopy (CAFM), supported by scanning tunneling microscopy (STM), thermodynamic modeling, and first-principles calculations. We focus on a prominent electronically active defect, $D^*$, which exhibits strong contrast over many unit cells. S-rich growth decreases the density of $D^*$ by a factor of 2.5, consistent with enhanced CrSBr phase stability and changes in gas-phase transport predicted by thermodynamic modeling. Lowering the absolute growth temperatures while maintaining the same temperature gradient further decreases the defect density by nearly an order of magnitude, consistent with decreased transport and growth rates. Combined experimental and theoretical analysis identifies a S-vacancy-related defect complex as the most likely origin of $D^*$. These results establish targeted growth optimization as an effective route to suppressing intrinsic defects and producing high-quality CrSBr for intrinsic-property measurements and deterministic defect engineering.


\begin{figure}
\scalebox{\figurescale}{\includegraphics[width=1\linewidth]{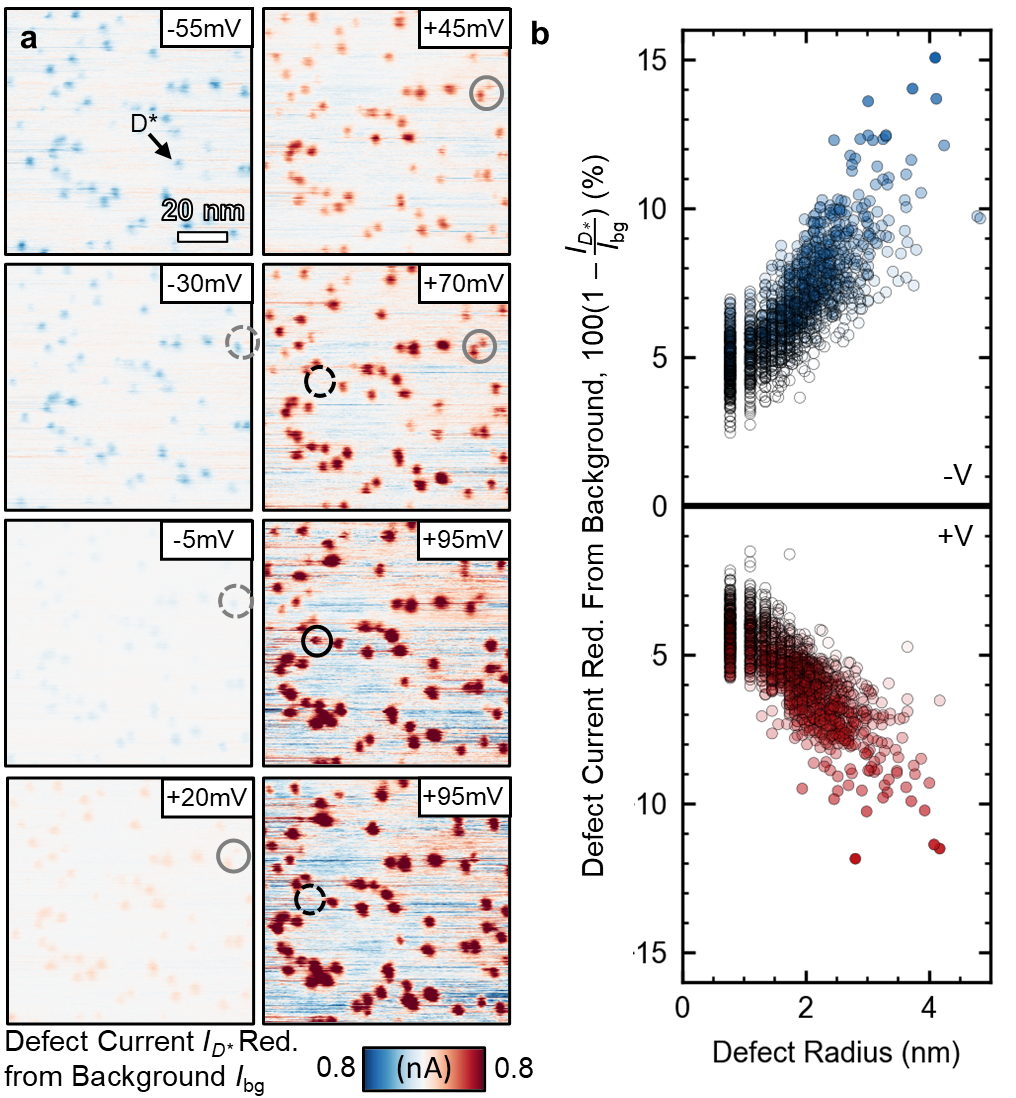}}
\caption{\textbf{Bias-dependent imaging and statistical analysis of the characteristic defect.}
\textbf{a}, Bias-dependent CAFM current maps acquired over a $100 \times 100$~nm$^2$ area at biases ranging from $-55$~mV to $+95$~mV. Circles mark defects that change during scanning. Dotted circles indicate the absence of a defect, while solid circles indicate its presence.
\textbf{b}, Defect radius as a function of current reduction, extracted from negative- and positive-bias CAFM images acquired over a $500 \times 500$~nm$^2$ area.}
\label{fig2}
\end{figure}

\textbf{Defects in CrSBr.} We synthesize CrSBr single crystals by CVT, as schematically illustrated in Fig.~\ref{fig1}c. For stoichiometric growth, high-purity Cr(s), S(s), and Br$_2$(l) are reacted in a 1:1:1 ratio inside a sealed, evacuated quartz ampoule. Crystal growth is performed in a horizontal two-zone furnace under a controlled temperature gradient ($\Delta T$), with source evaporation and crystallization zones maintained at $T_2=900$ and $T_1=800^\circ$C for approximately two weeks. After growth, the ampoule contains CrSBr single crystals as well as secondary phases, including CrBr$_3$, Cr$_2$S$_3$, and CrS, which are known to form under similar growth conditions~\cite{Song2025}. The resulting CrSBr crystals (Fig.~\ref{fig1}d) reach typical lateral dimensions of up to $5 \times 20$ mm and are handled under inert atmosphere.

We investigate the synthesized crystals using CAFM and scanning tunneling microscopy (STM). CAFM provides a robust, high-throughput method for quantifying defects in vdW materials under ambient conditions~\cite{Xu2023}, enabling statistical analysis over many samples and micrometer-scale areas. STM complements these measurements by providing atomic-resolution images of individual defects and confirming that the features observed by CAFM correspond to the same defect species. For CAFM, mechanically exfoliated CrSBr flakes are placed on an Au-coated p-type silicon substrate to ensure reliable electrical contact between the tip and sample (Fig.~\ref{fig1}e, see Methods, Supporting Information S1, and Supporting Fig. S1).

A representative optical micrograph of an exfoliated flake is shown in Fig.~\ref{fig1}f, and a corresponding large-area ($\SI{1} \times \SI{1}{\micro\meter}^2$) CAFM current map is shown in Fig.~\ref{fig1}g. The map reveals a characteristic defect, $D^*$, with a density of approximately $7 \times 10^{11}\si{\per\centi\meter\squared}$ for stoichiometric growth under an $800$-$900$$^\circ$C temperature profile. Higher-resolution CAFM images (Figs.~\ref{fig1}h,i) show that this defect exhibits pronounced electronic contrast and a spatial extent of several nanometers. Importantly, simultaneous acquisition of topographic information during CAFM scanning reveals no topographic changes associated with the current signal, proving that the signal does not arise from surface morphology and is consistent with an electronic defect (see Supporting Information S2 and Supporting Fig. S2). Moreover, we observe the identical defect in STM images (Figs.~\ref{fig1}j,k), further supporting our identification of $D^*$. 

2D Fast Fourier Transform analysis of CAFM images reveals an isotropic defect distribution (see Supporting Information S2 and Supporting Fig. S3). In rare instances, however, the defects exhibit 1D ordering along lines, suggesting the possible formation of dislocations within the material (see Supporting Information S2 and Supporting Fig. S4-5).

Notably, Br vacancies are challenging to resolve in CrSBr using STM~\cite{Klein.2022a,Telford.2020,Telford.2022}. Therefore, our analysis focuses exclusively on this prominent defect.

\textbf{Electronic role of the defect.} We continue by discussing the electronic signatures, voltage response, and statistical behavior of $D^*$. To this end, we collect CAFM images and apply a custom Python-based workflow that automatically analyzes images to identify and count defects (see Methods, Supporting Information S3, and Supporting Fig. S6). From these data, we extract relevant statistical metrics on an image-by-image basis.

We begin by studying the voltage-dependent defect response in CAFM measurements (Fig.~\ref{fig2}). We collect CAFM images over a fixed $100 \times 100~\si{\nano\meter\squared}$ area while varying the sample bias from $\SI{-55}{\milli\volt}$ to $\SI{+95}{\milli\volt}$ (Fig.~\ref{fig2}a). In the images, we show the absolute current response of the defect $I_{D^*}$ relative to the background current, $I_{\mathrm{bg}}$, i.e., the defect-free current values (see Methods section, Supporting Information S4, and Supporting Fig. S7). Interestingly, the defect exhibits a nearly symmetric response with respect to bias polarity around zero bias, characterized by a reduction of current magnitude at the defect site compared to the background. This behavior differs from typical electronic defects in layered materials, which show strongly asymmetric, bias-dependent contrast due to localized electronic states within the band gap~\cite{Schuler2019,Xu2023}. 

To statistically examine how bias polarity influences defect contrast, we acquire large field-of-view ($500 \times 500~\si{\nano\meter\squared}$) current maps at both positive and negative biases and analyze the current response for each defect in the images (see Methods section, Supporting Information S4, and Supporting Figs. S8-9). This is motivated by the observation of different defect sizes that we hypothesize may originate from identical defects located in different atomic planes within a layer, or from qualitatively similar defect complexes in which different numbers of atoms are missing. With our analysis, we statistically show the relationship between defect radius and relative current magnitude reduction (Fig.~\ref{fig2}b). The contrast is expressed as the percentage change in defect current $I_{D^*}$ relative to the background current $I_{\mathrm{bg}}$ of each image, which shows how the defect perturbs current flow as the tip scans across it. For both bias polarities, we consistently observe a symmetrical reduction in current magnitude relative to the background. By extracting both parameters using our workflow, we find an approximately linear dependence where larger defects produce stronger current suppression.

Assuming an ohmic tip-sample contact, the bias-symmetric reduction in CAFM current, together with the STM response, is consistent with the defect acting as a locally resistive electronic perturbation, suppressing charge transport and reducing the local density of states probed by STM in the investigated energy range. We further observe that a small subset of defect signatures emerge or disappear upon repeated scanning of the same area, as two examples are highlighted in (Fig.~\ref{fig2}a). The intermittent appearance of a subset of defects in repeated CAFM scans suggests that their electronic contrast is not purely structural, but may depend on a metastable charge state. Bias-induced trapping or detrapping at the defect could switch the defect between charge states with different local conductance, causing the defect to appear or disappear in the current maps. We note that this behavior changes the measured defect densities by less than 5\% (see Supporting Information S4 and Supporting Figs. S10-11) and therefore does not affect the statistical robustness of our analysis.

\begin{figure}
\scalebox{\figurescale}{\includegraphics[width=1\linewidth]{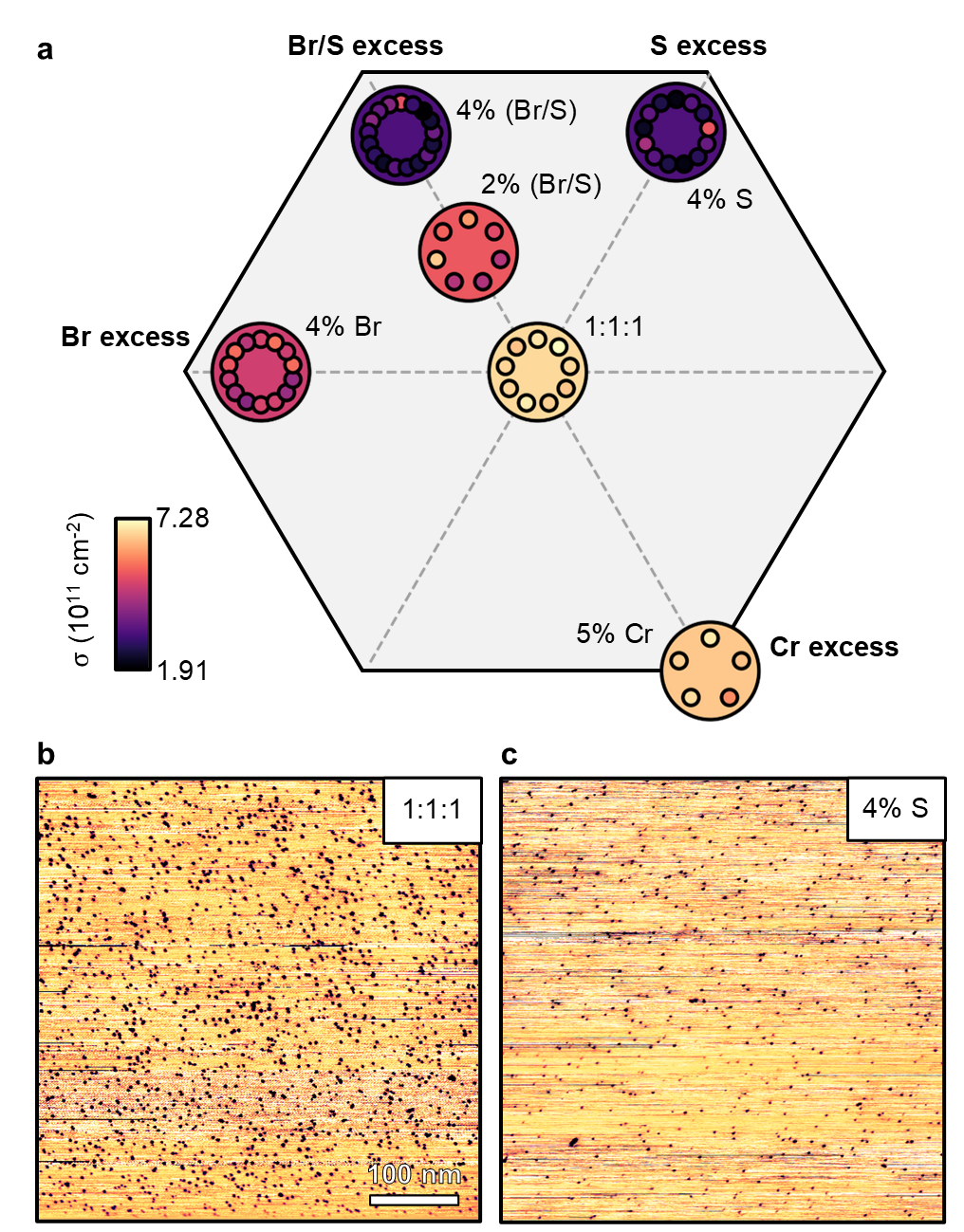}}
\caption{\textbf{Defect density as a function of elemental excess during CrSBr growth.}
\textbf{a}, Ternary growth-composition map showing defect densities extracted from CAFM measurements. All samples were grown using a temperature profile of $800$-$900^\circ\mathrm{C}$. Small circles represent individual measurements from $500 \times 500$~nm$^2$ CAFM images, while large circles indicate the mean defect density for each growth condition.
\textbf{b,c}, Representative CAFM current maps of CrSBr grown under stoichiometric conditions and with 4\% S excess, respectively.
}
\label{fig3}
\end{figure}

\textbf{Precursor stoichiometry control.} To study the effect of growth conditions on the concentration of defect $D^*$, we systematically vary the precursor stoichiometry by introducing controlled excess amounts of elemental Cr, S, or Br into the ampoule during growth. For the precursor control experiments, a fixed temperature gradient of $800$-$900\,^\circ\mathrm{C}$ is used during vapor transport. For each growth condition, single crystals of CrSBr are synthesized and bulk-like flakes are exfoliated onto Au/p$^{+}$-Si substrates. For each condition, we collect $500 \times 500$~nm$^2$ CAFM images on multiple exfoliated flakes, enabling statistically robust defect analysis. Fig.~\ref{fig3}a summarizes the results in the form of a ternary composition map. Moreover, a summary of the measured defect densities, determined by our Python workflow, and total measurement areas for all growth conditions is provided in Table~\ref{tab:measurement_summary}.

We observe the highest defect densities for stoichiometric growth, with $\sigma_{1:1:1}=(6.57 \pm 0.36) \times 10^{11}~\si{\per\centi\meter\squared}$, and a similar value for growth with 5\% Cr excess, $\sigma_{1.05:1:1}=(6.25 \pm 0.60) \times 10^{11}~\si{\per\centi\meter\squared}$. In contrast, increasing the S excess to 4\% leads to the lowest defect density observed for a single elemental excess, $\sigma_{1:1.04:1}=(2.66 \pm 0.71) \times 10^{11}~\si{\per\centi\meter\squared}$, corresponding to a reduction by a factor of about 2.5 compared to stoichiometric growth. A reduction in defect density is also observed for 4\% Br excess, although the effect is less pronounced than for S-rich growth. Increasing both S and Br excess to 4\% does not result in significantly more reduction than 4\% S only. From one representative images of each batch in Fig.~\ref{fig3}b,c, the nearest neighbor distance between defects decreases by over $d_{NN} = \SI{3}{\nano\meter}$ from $\SI{7.3}{\nano\meter}$ in the stoichiometric growth case to $\SI{10.6}{\nano\meter}$ in the 4\% S excess case.


\begin{table}[h!]
\centering
\resizebox{\columnwidth}{!}{%
\begin{tabular}{
    >{\raggedright\arraybackslash}p{3cm}
    p{3cm}
    p{3cm}
}
\toprule
Condition &
Density $\pm$ SD\newline
($10^{11}\,\mathrm{cm}^{-2}$) & Total area\newline ($\mu\mathrm{m}^{2}$) \\ 
\midrule
Stoichiometric & $6.57 \pm 0.36$ & 2.25 \\
5\% Cr         & $6.25 \pm 0.60$ & 1.25 \\
4\% S          & $2.66 \pm 0.71$ & 3.25 \\
4\% Br         & $4.05 \pm 0.56$ & 3.50 \\
2\% (Br/S)     & $4.54 \pm 1.00$ & 1.75 \\
4\% (Br/S)     & $2.65 \pm 0.58$ & 4.25 \\
\bottomrule
\end{tabular}%
}
\caption{Summary of defect densities and total CAFM measurement areas for different initial growth compositions using an $800$-$900^\circ\mathrm{C}$ temperature profile. Values following $\pm$ represent the sample standard deviation (SD).}
\label{tab:measurement_summary}
\end{table}

Moreover, we observe variations in defect concentrations between different exfoliated flakes grown under the same conditions, as well as spatial variations within individual exfoliated flakes (see Supporting Information S5 and Supporting Fig. S12). We attribute this spread to local variations in the growth environment within the ampoule. Spatial inhomogeneities in vapor composition and temperature along the growth direction can lead to position-dependent defect incorporation~\cite{Colombara2013,May2020}. These observations indicate that, while global defect densities can be statistically decreased through optimized growth conditions, local characterization remains important. In this context, CAFM provides a powerful tool for assessing spatial defect distributions at the nanoscale, which is particularly relevant for evaluating material quality for device applications.


\begin{figure}
\scalebox{\figurescale}{\includegraphics[width=1\linewidth]{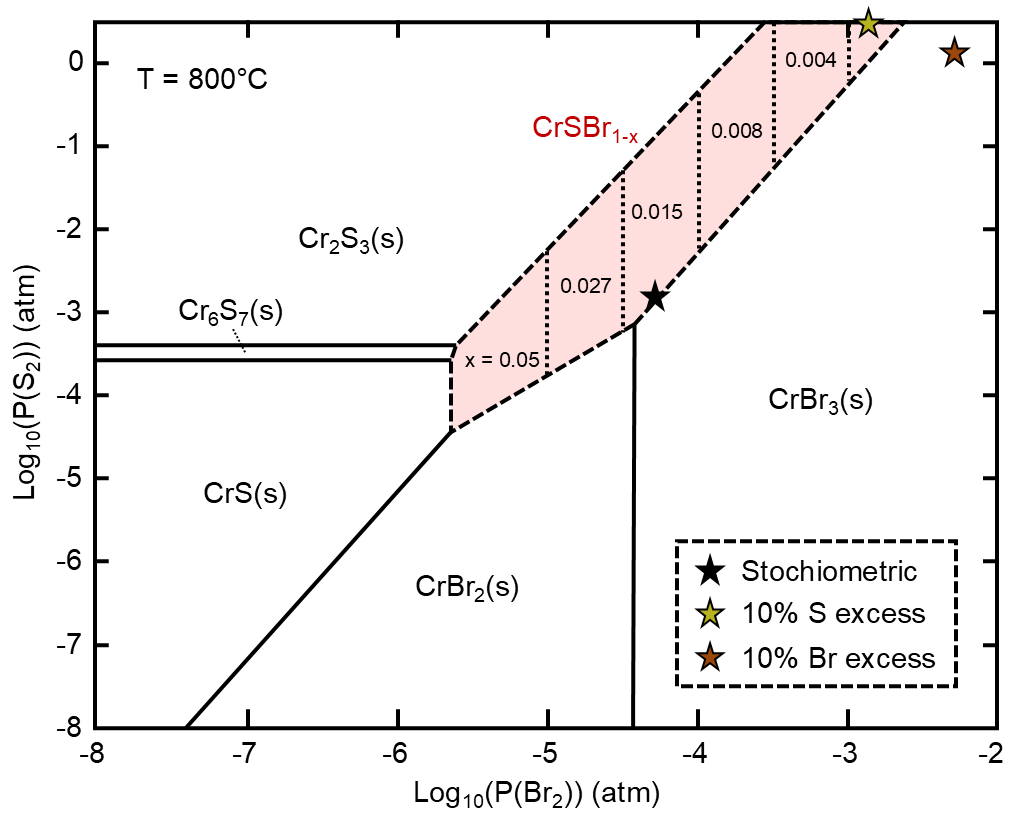}}
\caption{\textbf{Thermodynamic modeling of CrSBr growth.}
Calculated Kellogg-type phase stability diagram at $800^\circ\mathrm{C}$ showing the thermodynamically favored condensed phases as a function of the S and Br partial pressures. The $\star$ symbols mark the modeled growth conditions for the nominally stoichiometric composition and for compositions with 10\% S or 10\% Br excess. 
}

\label{fig4}
\end{figure}

\textbf{Thermodynamic modeling.} To gain insight into the decreased defect densities observed under S- and Br-rich growth conditions, we perform thermodynamic modeling of the CVT growth environment using the experimentally relevant synthesis parameters. We consider an ampoule composition corresponding to $0.25$~mol Cr, $0.25$~mol S, and $0.125$~mol Br$_2$ in a volume of $0.335$~L, assuming thermodynamic equilibrium at $900^\circ\mathrm{C}$ in the source evaporation zone and $800^\circ\mathrm{C}$ in the crystallization zone. 

The calculated gas phase consists primarily of S$_2$, CrBr$_4$, Br, SBr$_2$, Br$_2$, and S$_2$. At $800,^{\circ}$C, the corresponding partial pressures are $\log{10}(P{\mathrm{S}2}/\mathrm{bar})=-2.8$ and $\log{10}(P{\mathrm{Br}_2}/\mathrm{bar})=-4.3$. These conditions fall within the CrSBr-predominant region of the Kellogg-type diagram shown in Fig.~\ref{fig4}, which maps the thermodynamically favored condensed phases as a function of the S and Br partial pressures. This indicates that CrSBr is stable relative to the competing condensed phases considered in the model, although minor amounts of secondary phases still coexist. CrBr$_4$ is the only Cr-containing gas-phase species predicted by the model and is therefore identified as the dominant carrier for Cr transport during growth. 

We next examine the effect of modifying the initial stoichiometry. Increasing the S content by 10\% preserves the thermodynamic stability of CrSBr while simultaneously reducing the effective Br deficiency, as shown in Fig.~\ref{fig4}. In contrast, increasing the Br$_2$ content by 10\% shifts the calculated gas composition outside the CrSBr stability field. The Br-excess calculation should be interpreted cautiously, because added Br strongly changes the total pressure and gas speciation. Nevertheless, both S- and Br-rich calculations indicate a strong reduction in the concentration of the CrBr$_4$ transport species, with the strongest suppression found for the S-rich composition.

This reduction of CrBr$_4$ activity provides one plausible mechanism for the experimentally observed decrease in defect density under S- and Br-rich growth conditions. We therefore attribute the decreased defect incorporation not to a uniquely identified point-defect equilibrium, but to a combined change in phase stability, gas-phase chemistry, and decreased Cr transport flux.

\textbf{Temperature gradient control.} In addition to the thermodynamic controls discussed above, tuning growth kinetics through temperature is well known to control defect density~\cite{May2020,Sayers2020}. To this end, we also investigate how the temperature profile during CVT growth, and therefore, the transport flux affects the defect concentration. To isolate this effect, crystals are grown at 4\% S and Br excess while systematically lowering the absolute growth temperatures from $800$-$900^\circ\mathrm{C}$ to $750$-$850^\circ\mathrm{C}$ and $700$-$800^\circ\mathrm{C}$, while maintaining a constant end-to-end temperature difference of $\Delta T = \SI{100}{\celsius}$.


\begin{figure}
    \scalebox{\figurescale}{\includegraphics[width=1\linewidth]{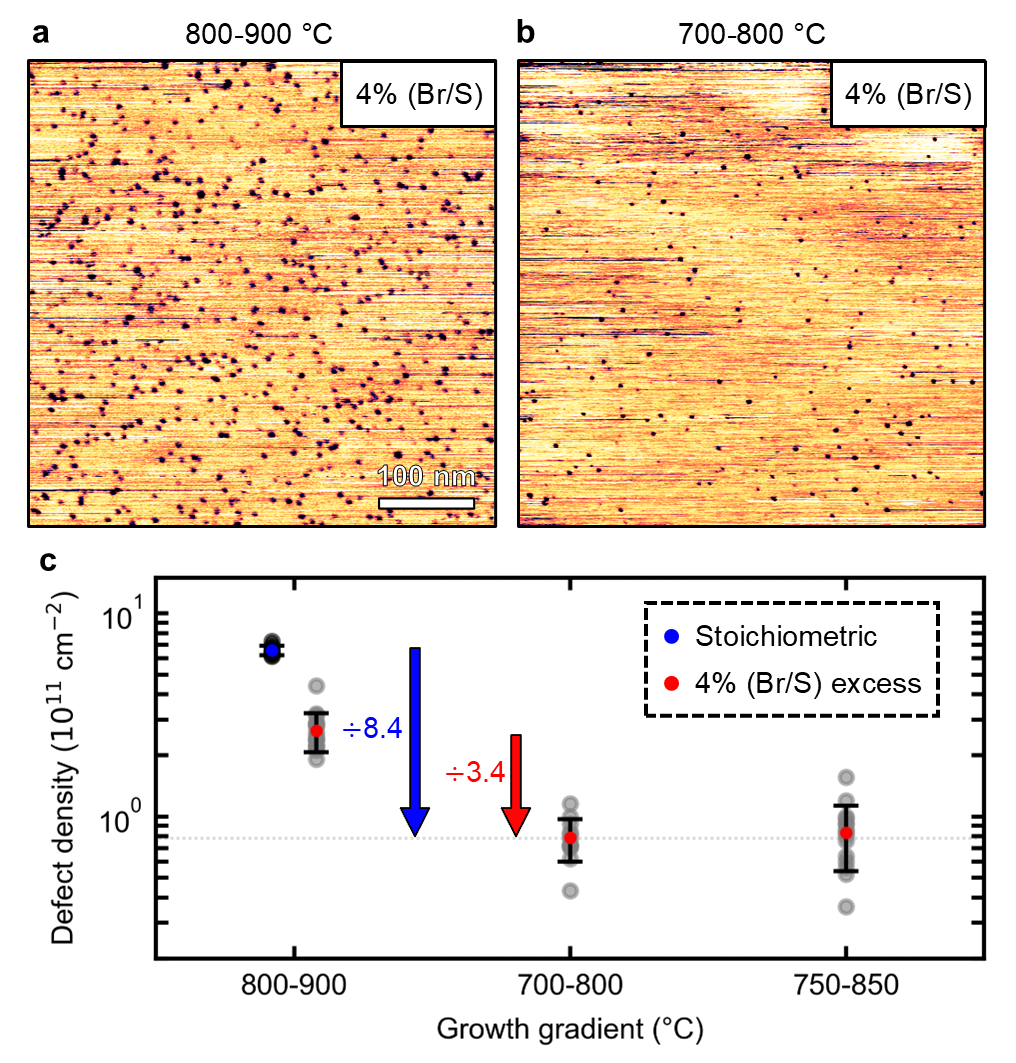}}
    \caption{\textbf{Effect of growth temperature gradient on defect density.}
        \textbf{a,b}, Representative CAFM current maps of CrSBr grown with 4\% Br/S excess at temperature profiles of $800$-$900^\circ\mathrm{C}$ and $700$-$800^\circ\mathrm{C}$, respectively.
        \textbf{c}, Statistical comparison of defect densities for three different growth temperature profiles: $800$-$900^\circ\mathrm{C}$, $700$-$800^\circ\mathrm{C}$, and $750$-$850^\circ\mathrm{C}$.
	}
    \label{fig5}
\end{figure}

Figs.~\ref{fig5}a and \ref{fig5}b show representative CAFM images obtained using the highest and lowest temperature profiles, respectively. A striking decrease in defect density is observed for growth at lower absolute temperatures. The statistical analysis is summarized in Fig.~\ref{fig5}c and Table~\ref{tab:temperature_summary}. For growth with 4\% Br/S excess at $700$-$800^\circ\mathrm{C}$, we obtain a defect density of $\sigma^{700-800^\circ\mathrm{C}}_{1:1.04:1.04}=(0.79 \pm 0.19)\times10^{11}\si{\per\centi\meter\squared}$, corresponding to an almost order-of-magnitude decrease compared with standard stoichiometric growth at $800$-$900^\circ\mathrm{C}$. From a representative image acquired under the $700$-$800^\circ\mathrm{C}$ condition, the mean nearest-neighbor distance is calculated as $d_{\mathrm{NN}}=\SI{18.8}{\nano\meter}$. This is approximately 1.8 times larger than for growth with 4\% Br/S excess at $800$-$900^\circ\mathrm{C}$ ($\SI{10.7}{\nano\meter}$) and 2.6 times larger than for stoichiometric growth at $800$-$900^\circ\mathrm{C}$ ($\SI{7.3}{\nano\meter}$). Growth with 4\% Br/S excess at $750$-$850^\circ\mathrm{C}$ yields a similar defect density of $\sigma^{750-850^\circ\mathrm{C}}_{1:1.04:1.04}=(0.84 \pm 0.30)\times10^{11}\si{\per\centi\meter\squared}$. The lowest defect density observed in a single $500 \times 500$~nm$^2$ image is $0.36\times10^{11}\si{\per\centi\meter\squared}$, a decrease by a factor of approximately 18 compared with standard stoichiometric growth. The distribution observed for both cases is likely attributable to spatial inhomogeneity within individual crystals and the sample-to-sample variation discussed above (see Supporting Information S5 and Supporting Figs. S12-13).

\begin{table}[h!]
\centering
\begin{tabular}{
    >{\raggedright\arraybackslash}p{3.3cm}
    p{3cm}
    p{2cm}
}
\toprule
Temperature profile &
Density $\pm$ SD\newline
($10^{11}\,\mathrm{cm}^{-2}$) &
Total area\newline
($\mu\mathrm{m}^{2}$) \\
\midrule
$750$-$850^\circ\mathrm{C}$                     & $0.84 \pm 0.30$ & 3.75 \\
$700$-$800^\circ\mathrm{C}$                     & $0.79 \pm 0.19$ & 3.00 \\
\bottomrule
\end{tabular}
\caption{Summary of defect densities and total CAFM measurement areas for samples grown with a 4\% Br/S excess using different temperature profiles.}
\label{tab:temperature_summary}
\end{table}

Although the temperature difference was kept constant at $\Delta T = 100^\circ\mathrm{C}$, changing the absolute growth temperatures strongly affected the defect density.
In CVT growth, the absolute temperature profile determines the vapor pressures, transport-species fluxes, and degree of supersaturation at the crystallization front. Lowering the overall temperature range decreases the transport and growth rates, such as for CrBr$_4$ in the gas phase, which is accompanied by a lower defect density under the conditions studied here.
This effect is complementary to the stoichiometry trends discussed above. While S- and Br-rich conditions modify the thermodynamic growth environment and gas-phase chemistry, the temperature profile tunes transport rates and growth kinetics.

To examine how the lower-temperature profile affects phase stability, we calculated equilibria for source compositions with Cr:S:Br ratios of 1:1:1 and with 4\% excess S and/or Br. Source-zone equilibria were evaluated at $900^\circ\mathrm{C}$ and $800^\circ\mathrm{C}$ at constant volume, while target-zone equilibria were evaluated at $800^\circ\mathrm{C}$ and $700^\circ\mathrm{C}$ at constant pressure, using the pressure and gas-phase composition obtained from the corresponding source calculation. CrSBr formation in the colder zone is predicted only for three conditions: the stoichiometric 1:1:1 composition at $800^\circ\mathrm{C}$, and the S-rich 1:1.04:1 composition at both $800^\circ\mathrm{C}$ and $700^\circ\mathrm{C}$. In contrast, any Br excess leads to CrBr$_3$ formation. These results indicate that moderate S excess stabilizes CrSBr formation at both target-zone temperatures, whereas Br excess drives the system toward competing chromium bromide phases.


\begin{figure}
\scalebox{\figurescale}{\includegraphics[width=1\linewidth]{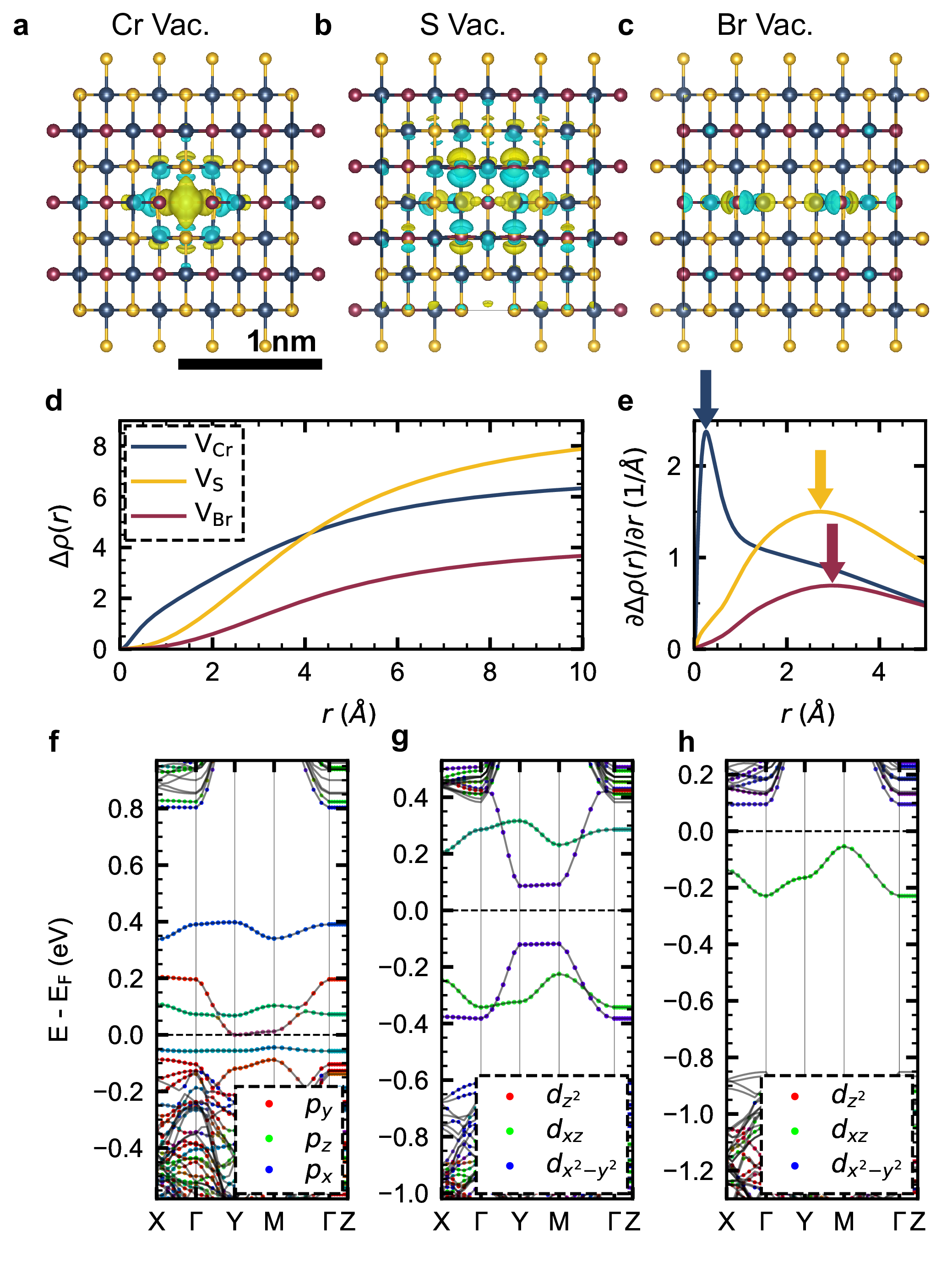}}
\caption{\textbf{Spatial extent of charge-density perturbations from point vacancies in CrSBr.}
    \textbf{a-c}, Charge-density perturbations induced by Cr, S, and Br vacancies, shown at the same isosurface value. Scale bar, 1~nm.
    \textbf{d}, Integrated charge-density perturbation $\Delta \rho(r)$ as a function of radius $r$ around the vacancy center for the vacancies shown in \textbf{a-c}.
    \textbf{e}, Radial derivative $\partial \Delta \rho(r)/\partial r$ for the corresponding vacancies with arrows indicating the maximum values.
    \textbf{f-h}, DFT+$U$ band structures of V$_{Cr}$, V$_{S}$, and V$_{Br}$ including orbital projections on nearest-neighbor sites.
}
\label{fig6}
\end{figure}

\textbf{Microscopic structure.} We next investigate the possible microscopic structure of the defect using density functional theory (DFT) and identify an S-vacancy-related structure as a potential origin of $D^*$. To compare the perturbations induced by different point defects, we calculate the local lattice distortion (see Supporting Information S6 and Supporting Fig. S14), charge-density redistribution, and projected band structures for individual vacancy configurations. The aim of this analysis is to quantify the strength, the spatial extent, and the electronic footprint of the perturbation associated with each vacancy type. Fig.~\ref{fig6}a-c shows the real-space response for single Cr, S, and Br vacancies, together with the integrated charge-density modulation as a function of the radius $r$ around the vacancy center. Among the three cases, the S vacancy produces the most extended charge-density perturbation, whereas the perturbations associated with Cr and Br vacancies are more localized. The extended perturbation is in direct connection with our CAFM results that show the defect extent over nm length scales (see Fig.~\ref{fig1}h,i).

We quantify this behavior by evaluating $\Delta \rho(r) = \sum_{r'<r}|\rho_\text{pri}(r') - \rho_\text{vac}(r') - \rho_\text{at}(r')|/\rho_{\text{vac, total}}$, representing the charge densities $\rho(r)$ in units of $\text{\AA}^{-3}$ of a pristine CrSBr supercell, a CrSBr supercell with the vacancy atom, and the single (vacancy) atom, normalized by the total charge of the supercell with the vacancy, together with its radial derivative $\partial \Delta \rho(r)/\partial r$ (Fig.~\ref{fig6}d,e). The magnitude of $\Delta \rho(r=10\text{\AA})$ is largest for the S vacancy, reaching approximately 8, compared with about 6 for the Cr vacancy and about 4 for the Br vacancy. The derivative further shows that the charge-density perturbation of the Cr vacancy is dominated by changes in the immediate vicinity of the vacancy, whereas the S vacancy perturbs the charge density over a larger radius. The Br vacancy also gives rise to a more spatially extended modulation, but this response is weaker overall and primarily anisotropic along the $a$ direction. The perturbed electronic structures, depicted in Fig.~\ref{fig6}f-h, further show that the single Cr vacancy has a tendency to pin the chemical potential to the valance band maximum, while the S and Br vacancies pin it more likely to the conduction band minimum. Among the three candidates, the S vacancy imprints the most symmetric impurity electronic structure around the Fermi level, with occupied and unoccupied states of mostly Cr d$_{xz}$ and d$_{x^2y^2}$ character at the neighboring Cr atoms. This behavior connects to our CAFM result of the bias-symmetric defect behavior(see Fig.~\ref{fig2}a).

Taken together, the experimental observations and theoretical calculations suggest that $D^*$ is most consistent with a S-related vacancy complex rather than an isolated Cr or Br vacancy. This assignment is supported by the strong decrease of defect density under S-rich growth, the stabilization of CrSBr and suppression of CrBr$_4$ transport under S excess, the extended charge-density perturbation predicted for S vacancies by DFT, and the imprinted rather symmetric impurity electronic structure around the Fermi level. At the same time, the defect cannot be assigned to a uniquely identified isolated point defect based on the present data alone. The nearly order-of-magnitude decrease in $D^*$ density achieved by lowering the absolute growth temperatures further shows that defect incorporation is strongly affected by transport kinetics. Thus, the formation of $D^*$ is likely governed by both local defect chemistry and the kinetic conditions during CVT growth.

\textbf{Conclusion.} In summary, we identify the characteristic defect $D^*$ as likely related to an S-vacancy and demonstrate that its density can be substantially decreased through targeted optimization of CrSBr growth conditions. A moderate excess of elemental S significantly decreases the defect density, while lowering the absolute growth temperatures at a constant temperature gradient leads to a further reduction of nearly an order of magnitude for growth with 4\% Br/S excess. These results support that defect incorporation in CrSBr depends not only on the starting composition and temperature gradient, but also on the absolute growth temperatures, transport rates, and growth kinetics.

The dominant defect observed by CAFM produces a locally resistive electronic perturbation. Its bias-symmetric current suppression and STM response are consistent with decreased local conductance and a modified local density of states in the probed energy range. Combined with the DFT calculations, these observations support the assignment of $D^*$ to an S-vacancy-related complex, although alternative structures involving Br vacancies or Cr interstitials cannot be fully excluded. Unambiguous identification will require additional atomic-scale measurements, such as low-temperature bias-dependent STM.

Together, these results establish growth optimization as an effective route toward improving CrSBr crystal quality. In particular, combining modified starting compositions with lower growth temperatures provides a practical strategy for suppressing native defects. Further optimization of the transport conditions and temperature profile will be important for accessing the intrinsic electronic and optical properties of CrSBr and providing a well-controlled host for deterministic defect engineering~\cite{Klein2026}.

%
%
\section{Acknowledgements}
S.R.T., F.M.R., and J.K. acknowledge support from the National Science Foundation under Trailblazer Engineering Impact Award No. 2421694 for work related to the theoretical modeling of defects. F.M.R. and J.K. acknowledge support from the U.S. Department of Energy, Office of Science, Office of Basic Energy Sciences, Division of Materials Sciences and Engineering under Award No. DE-SC0025387 for the experimental investigation and analysis of defect structures. Z.S. and I. P. were supported by project LUAUS25268 from Ministry of Education, Youth and Sports (MEYS). Z.S. and D.S. were supported by the Advanced Multiscale Materials for Key Enabling Technologies project, supported by the Ministry of Education, Youth, and Sports of the Czech Republic. Project No. CZ.02.01.01/00/22\_008/0004558, Co-funded by the European Union. The authors acknowledge the MIT SuperCloud and Lincoln Laboratory Supercomputing Center for providing HPC and database resources~\cite{reuther2018interactive}. The authors acknowledge the MIT Office of Research Computing and Data for providing high performance computing resources that have contributed to the research results reported within this paper. The authors acknowledge the MIT Office of Research Computing and Data for providing high performance computing resources that have contributed to the research results reported within this paper. This work was performed in part in the MIT.nano Characterization Facilities on the Cypher VRS enabled by DURIP award (N000142012203). We acknowledge use of the NanoMechanical Technology Laboratory in the Department of Materials Science and Engineering at MIT. M.R., S.G., and S.v.H. acknowledge support from the Vidi ENW research programme of the Dutch Research Council (NWO) under Grant DOI: 10.61686/YDRHT18202 (file number VI.Vidi.233.077). The ab initio calculations have been performed on the Dutch National Supercomputer Snellius.

\section{Author contributions}
J.K. conceptualized and supervised the project and together with S.R.T. designed the experiments. Z.S. and I. P. synthesized bulk crystals of CrSBr. S.R.T. fabricated samples. S.R.T. collected CAFM data. R.D. collected STM data. D.S. performed thermodynamic modeling of growth. S.G., S. von H. and M.R. provided ab-initio DFT calculations. S.R.T. and J.K. analyzed the data. All authors discussed the results. The manuscript was written by J.K. and S.R.T. with input from all co-authors.

%

\section{Methods}

\subsection*{Crystal growth}

Bulk CrSBr single crystals were synthesized by chemical vapor transport (CVT) in evacuated and sealed quartz ampoules. For the standard growth conditions, elemental Cr, S, and Br$_2$ were loaded in a nominally stoichiometric molar ratio of 1:1:1 and reacted under a temperature gradient of $800$-$900^\circ\mathrm{C}$ for several days, yielding plate-like single crystals. Additional growths were performed under otherwise identical conditions while introducing controlled excess amounts (up to 5\%) of elemental Cr, S, or Br$_2$ into the ampoule.

Further growths were carried out using a 4\% excess of Br and S with decreased absolute temperatures while maintaining the same end-to-end temperature difference ($\Delta T = 100^\circ\mathrm{C}$). Specifically, crystals were grown under temperature gradients of $750$-$850^\circ\mathrm{C}$ and $700$-$800^\circ\mathrm{C}$. All resulting crystals were handled and stored under inert conditions prior to characterization. Crystals grown at lower temperature gradient were smaller in size (1mm diameter and under 5mm length), attributed to the decrease of CrBr$_4$ in the gas phase from the 4\% Br/S excess and lower temperature.

\subsection*{Conductive atomic force microscopy}

Two different sample geometries were prepared for CAFM measurements. Most samples were fabricated by mechanical exfoliation of CrSBr crystals onto highly boron-doped p$^{+}$ Si substrates. To improve electrical grounding and minimize the influence of native oxide layers, most substrates were metallized prior to exfoliation by sputtering approximately 10 nm of Au or Au/Pd. CrSBr flakes were then mechanically exfoliated directly onto the metal-coated surface. The thickness of exfoliated flakes was verified using atomic force microscopy.

In addition to the above, selected measurements were performed on mm-thick bulk CrSBr crystals. In this case, small CVT-grown crystals were directly mounted onto a sample puck using conductive epoxy.

Samples prepared using the mechanical exfoliation method consistently exhibited improved signal during CAFM measurements, enabling more reliable current mapping. More details about the sample preparation can be found in Supporting Information S1 and Supporting Fig. S1.

Measurements were made using Cypher VRS and Cypher ES atomic force microscopes equipped with single- or dual-gain current amplifiers. CAFM measurements were typically conducted using sample biases between $-500$~mV and $+500$~mV, scan rates of 2-20~Hz, and contact-mode deflection setpoints between 0 and 200~mV. No systematic differences in defect density were observed when using higher setpoints or bias voltages outside this range. Therefore, imaging conditions were adjusted to optimize signal quality. Measurements were performed using conductive Pt-/PtIr-coated or diamond tips, including BudgetSensors ElectriMulti75E-G, BudgetSensors All-in-One-DD, and Bruker SCM-PlT probes. No significant differences in defect appearance or defect density were observed between the different tip types. Image resolution for defect counting images was set between 0.5 nm and 1 nm per pixel. Where possible, both trace and retrace current maps were acquired and cross-checked for most accurate defect counting. 

\subsection*{CAFM image Python analysis and visualization}

CAFM images were analyzed using a custom-built Python workflow. For typical defect counting in Python, line-by-line median correction was applied to the raw data to make the defect appearance clearer. A Gaussian blur and/or normalization was then applied to improve defect visibility further for reliable counting. Images were thresholded and defects were detected using the OpenCV package; misidentifications were filtered out on the basis of size and aspect ratio. Manual counting was employed for certain images, especially those with significant artifacts such as scan lines, drift, non-uniform tip shape, and changing sample-tip contact. A subset of images was manually counted and put through the workflow to validate the results from the workflow. From this result, we estimate the workflow error (in misidentifying defect density) to be 15\% at maximum. The workflow generally underestimates the density as the contour-finding algorithm considers all touching contours to be one contour, thus systematically reducing the density of defects, which is particularly relevant for images with high defect density, where defects appear closer together in $500 \times 500$~nm$^2$ view. Error is lower for images with lower density since the defects are farther apart and thus more easily detectable. More details can be found in Supporting Information S3 and Supporting Fig. S6.

For the bias series shown in Fig.~\ref{fig2}a, a line-by-line median correction was applied, followed by a polynomial background subtraction, followed by another mean background subtraction, to isolate the defects of interest and set the background value at 0 nA. Further details are provided in Supporting Information S4 and Supporting Fig. S7. More details about the defect current reduction calculation shown in Fig.~\ref{fig2}b are also detailed in Supporting Information S4 and Supporting Fig. S8-9. 

For best visualization, all CAFM images in the manuscript and Supporting Information are displayed with a line-by-line median correction and 3$^{rd}$ order polynomial background subtraction unless otherwise noted. Positive bias CAFM images are displayed with dark defects and negative bias CAFM images are displayed with bright defects.

\subsection{Scanning tunneling microscopy}

Topographic images were acquired at room temperature using a Unisoku UHV-LT four-probe scanning tunneling microscope operated with a Nanonis controller and integrated with a scanning electron microscope for precise positioning of the scan area. Bulk CrSBr crystals were cleaved \emph{in situ} under ultrahigh vacuum to expose clean surfaces prior to measurement.

For best visualization, all STM images in the manuscript and Supporting Information are displayed with a line-by-line median correction and 3$^{rd}$ order polynomial background subtraction unless otherwise noted. 

\subsection{Ab-initio calculations}

First-principles calculations were performed within density functional theory (DFT) using the Vienna Ab Initio Simulation Package (VASP)~\cite{KRESSE199615, PhysRevB.54.11169} on the Dutch National Supercomputer Snellius.
Electronic exchange and correlation were treated within the generalized gradient approximation (GGA) using the Perdew-Burke-Ernzerhof (PBE) functional~\cite{PhysRevLett.77.3865}. The on-site Coulomb interaction on the Cr 3d orbitals was described using the rotationally invariant GGA+U approach with U=2.5 eV and J=0.4 eV. These parameters were obtained from constrained random phase approximation calculations for single-layer CrSBr in an effective dielectric environment with $\varepsilon_{\mathrm{env}} \approx 8$~\cite{rudenko_dielectric_2023}. Defective structures were modeled using $4 \times 3$ supercells with bilayers in the $z$-direction separated by 14 \AA\ of vacuum. Brillouin-zone integrations were carried out using Monkhorst-Pack $3 \times 3 \times 1$ k-point meshes, and a plane-wave kinetic-energy cutoff of $500$ eV was employed. All atomic positions were fully relaxed until the residual Hellmann-Feynman forces on every atom were below 0.01~$\mathrm{eV \text{\AA}^{-1}}$.

\subsection{Thermodynamic growth modeling}

Phase equilibria in the Cr-S-Br system were modeled using FactSage thermodynamic database and calculation software~\cite{Bale2016}. Gibbs energy functions of stoichiometric phases and gas phase species were based on standard enthalpies of formation and 
entropies referred to T = 298.15 K together with the temperature dependence of heat capacity. Thermodynamic data of pure binary sulfides and bromides and gas 
phase constituents were taken from the FactPSBase database. Enthalpy of formation DHf = -275.0 kJ mol$^{-1}$ of CrSBr and the respective Br and S vacancy formation enthalpies 186.4 and 320.8 kJ mol$^{-1}$ were calculated using MedeA VASP software package (PAW basis, PBE exchange correlation functional, PW cut-off energy 400 eV).
The entropy and the temperature dependence of heat capacity of CrSBr were approximated by Neumann-Kopp additive rule based on solid CrCl$_3$ and Cr$_2$S$_3$ virtual constituents. The Kellogg phase diagrams were mapped in Phase Diagram module while the CVT process was modelled in Equilibrium module of the FactSage software.

%
%

\bibliographystyle{naturemag}
\bibliography{full}

\end{document}


\setcounter{figure}{0} 
\renewcommand{\thefigure}{S\arabic{figure}}

\renewcommand{\thetable}{S\arabic{table}}

\fontsize{11pt}{13pt}\selectfont

\makeatletter
\renewcommand\@make@capt@title[2]{%
    \@ifx@empty\float@link{\@firstofone}{\expandafter\href\expandafter{\float@link}}%
    \fontsize{11pt}{13pt}\selectfont\textbf{#1}\@caption@fignum@sep#2
}
\renewcommand\figurename{FIG.}
\makeatother

\title{Supporting Information:\\ Growth-controlled suppression of electrically active defects in CrSBr}
%
%
\author{Sara~R.~Tulchinsky}
\affiliation{Department of Materials Science and Engineering, Massachusetts Institute of Technology, Cambridge, Massachusetts 02139, USA}
\affiliation{Department of Physics and Astronomy, Wellesley College, Wellesley, Massachusetts 02481, USA}
\author{Sergii~Grytsiuk}
\affiliation{Faculty of Physics, Bielefeld University, 33501 Bielefeld, Germany}
\affiliation{Institute for Molecules and Materials, Radboud University, Heijendaalseweg 135, 6525AJ Nijmegen, The Netherlands}
\author{Shen~van~Hassel}
\affiliation{Faculty of Physics, Bielefeld University, 33501 Bielefeld, Germany}
\affiliation{Institute for Molecules and Materials, Radboud University, Heijendaalseweg 135, 6525AJ Nijmegen, The Netherlands}
\author{Iva~Plutnarová}
\affiliation{Department of Inorganic Chemistry, University of Chemistry and Technology Prague, Technická 5, 166 28 Prague 6, Czech Republic}
\author{Rami~Dana}
\affiliation{Department of Materials Science and Engineering, Massachusetts Institute of Technology, Cambridge, Massachusetts 02139, USA}
\author{David~Sedmidubský}
\affiliation{Department of Inorganic Chemistry, University of Chemistry and Technology Prague, Technická 5, 166 28 Prague 6, Czech Republic}
\author{Zdenek~Sofer}
\affiliation{Department of Inorganic Chemistry, University of Chemistry and Technology Prague, Technická 5, 166 28 Prague 6, Czech Republic}
\author{Malte~Rösner}
\affiliation{Faculty of Physics, Bielefeld University, 33501 Bielefeld, Germany}
\affiliation{Institute for Molecules and Materials, Radboud University, Heijendaalseweg 135, 6525AJ Nijmegen, The Netherlands}
\author{Frances~M.~Ross}
\affiliation{Department of Materials Science and Engineering, Massachusetts Institute of Technology, Cambridge, Massachusetts 02139, USA}
\author{Julian~Klein}\email{jpklein@mit.edu}
\affiliation{Department of Materials Science and Engineering, Massachusetts Institute of Technology, Cambridge, Massachusetts 02139, USA}
%
%

%
\maketitle
%
%

\tableofcontents

\section{S1: CAFM Sample Preparation Mechanisms}

\begin{figure*}
\scalebox{\figurescale}{\includegraphics[width=1\linewidth]{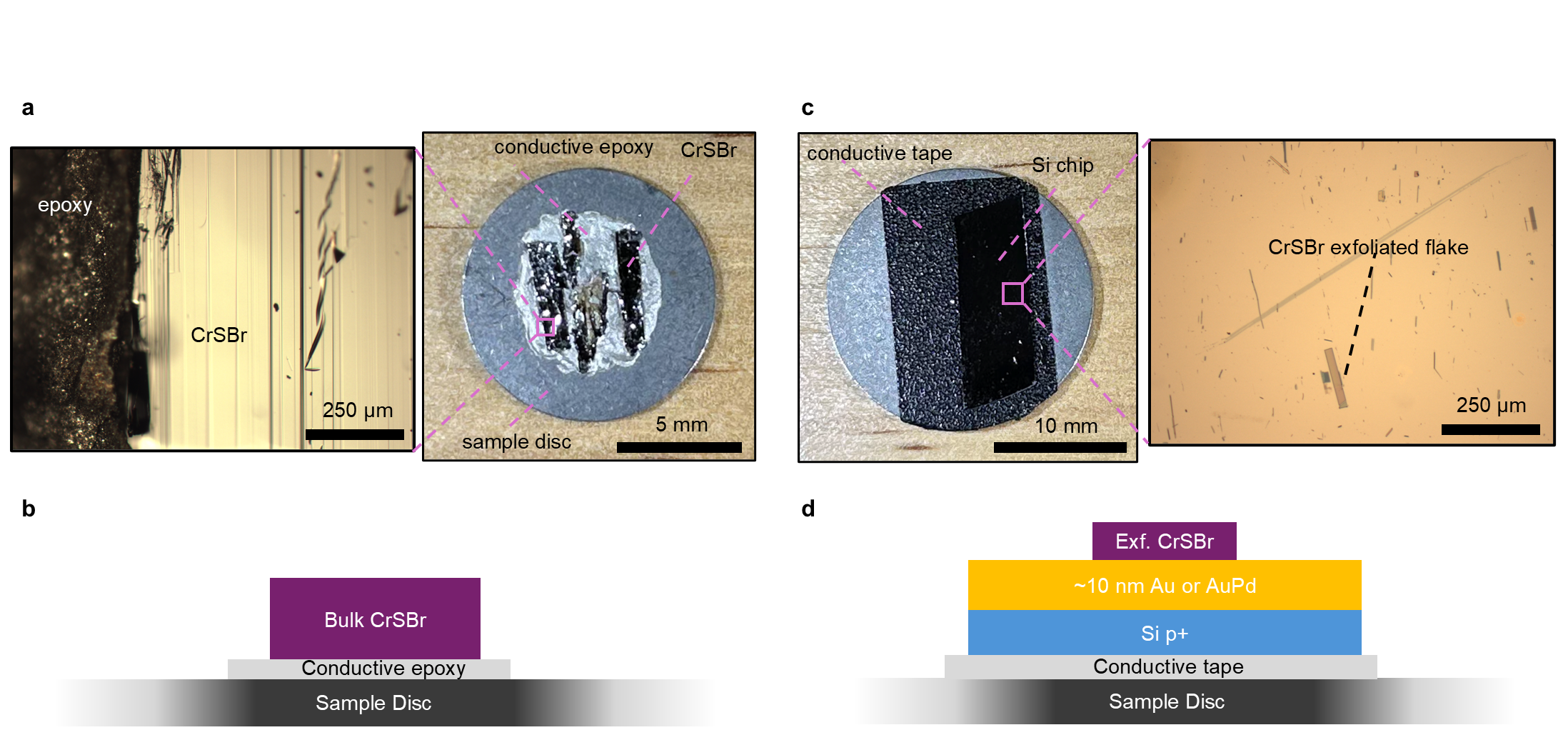}}
\caption{
\textbf{Optical micrographs and schematic of sample preparation mechanisms for CAFM.}
~\textbf{a}, Optical micrograph of a typical bulk crystal sample with mm-thick CrSBr flakes glued onto conductive epoxy. A zoom in of the mm-thick CrSBr flakes are shown on the left. ~\textbf{b}, Schematic stacking sequence of bulk CrSBr sample. ~\textbf{c}, Optical micrograph of typical ‘2D-like sample’ with exfoliated CrSBr flakes atop a sputtered p$^{+}$ wafer, attached using conductive carbon tape. Exfoliated flakes are shown on the zoom in. ~\textbf{d}, Schematic stacking sequence of 2D-like CrSBr sample.
}
\label{fig:sample preps}
\end{figure*}

Two different sample preparation methods were used to collect CAFM data. First, bulk CVT CrSBr flakes were glued directly onto a stainless steel AFM specimen discs (Ted Pella) using two-parts conductive epoxy (Ted Pella) cured at low temperatures (<100$^{\circ}$ C) to avoid sample degradation in a vacuum oven. Flat crystals with mm dimensions were selected, and the top layer was exfoliated with scotch tape directly before measurement to expose a fresh surface. This sample fabrication is demonstrated in Fig.~\ref{fig:sample preps}a,b. 

The sample preparation was further optimized for stronger sample-tip contact to the fabrication shown in Fig.~\ref{fig:sample preps}c,d. Boron-doped (500 - 550 $\si{\micro\meter}$ thick) Si wafers (University Wafer, 0.001-0.005 $\Omega$-cm resistivity) were sputter-coated with approximately 10 nm of Au or Au-Pd to negate any effects of the native oxide on the Si. Then CrSBr flakes were exfoliated onto the surface using typical 2D materials exfoliation. The flakes were typically 10-200 nm thick, verified by tapping mode AFM. We found that this sample preparation method was the most reliable for sample-tip contact and thus was used for most images included in this paper.

\section{S2: Nature of Defects}

\subsection{A. Topographic Confirmation of Electronic Defects}

To ensure that the observed defects are electronic in nature, instead of caused by surface morphology, we investigate if topography is correlated with the defect signature we observe. To this end, we collect and correlate topographic images that are collected simultaneously during CAFM image acquisition. 

\begin{figure}
    \scalebox{\figurescale}{\includegraphics[width=1\linewidth]{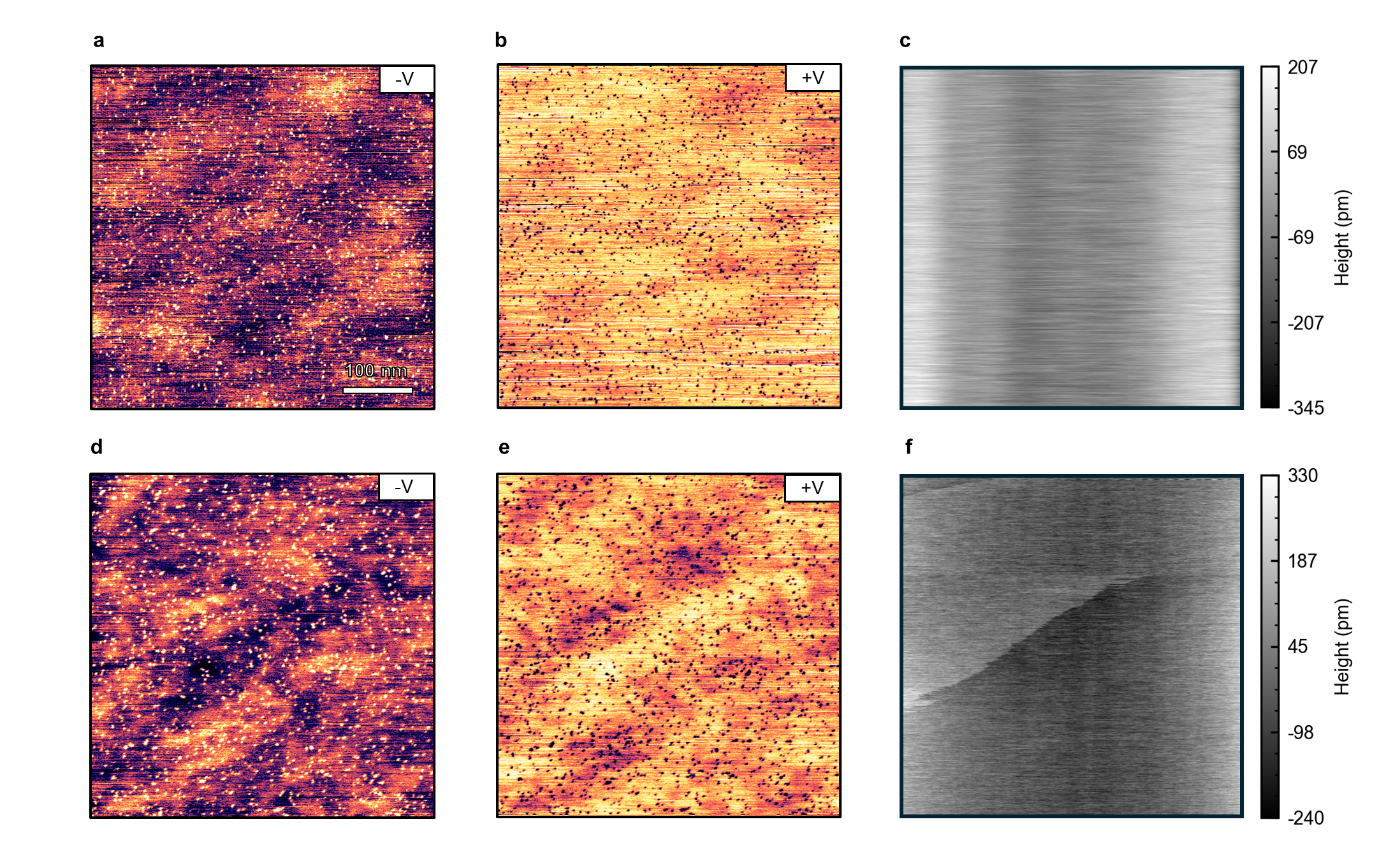}}
    \caption{\textbf{Topographic confirmation of electronic defect nature.}
        ~\textbf{a},~\textbf{b}, -V, +V bias flip $500 \times 500$~nm$^2$ image pair (stoichiometric growth batch at $800$-$900^\circ\mathrm{C}$) with ~\textbf{c}, corresponding topography image. ~\textbf{d},~\textbf{e}, -V, +V bias flip $500 \times 500$~nm$^2$ image pair (stoichiometric growth batch at $800$-$900^\circ\mathrm{C}$) with corresponding topography image in ~\textbf{f}.
	}
    \label{topography_fig}
\end{figure}

Fig.~\ref{topography_fig} demonstrates two examples of $500 \times 500$~nm$^2$ CAFM images that were collected for a negative and a positive bias of the same area, respectively, along with the corresponding topography image from the positive bias case. These images are from the stoichiometric growth batch at $800$-$900^\circ\mathrm{C}$. A topographic image is taken with both the positive and negative bias image, but we show just the one taken with the positive bias image. We do not observe a significant difference between them. The topographic images are presented with first-order flattening correction applied, which removes scanner/sample tilt and other artifacts, allowing us to see a more quantitatively accurate version of the sample topography. For the first case, the CAFM images are in Fig.~\ref{topography_fig}a,b while the topographic image is in Fig.~\ref{topography_fig}c. In the second case, the CAFM images are in Fig.~\ref{topography_fig}d,e with the topographic image in Fig.~\ref{topography_fig}f. For both cases, the topography image is atomically-flat with a height range within 600 pm (6 $\text{\AA}$), and the RMS values for roughness are 0.69 $\text{\AA}$ and 0.78 $\text{\AA}$, respectively, much less than the thickness of a single layer of CrSBr with 79 $\text{\AA}$~\cite{Beck.1990}. For both cases, in the topographic image, typical AFM ‘bow’ artifact is observed which originates from the cantilever motion. In the second example Fig.~\ref{topography_fig}d-f, an atomic step is imaged, whose signature is partly visible in the current retrace images. This type of feature is discussed in Fig.~\ref{steps}. 

In all cases, we find that the surface is atomically flat in the topographic image despite the presence of defects in both bias conditions in the CAFM images. This suggests that the observed defects are indeed electronic in nature. 

\subsection{B. Defect Spatial Properties in Varied CrSBr}

\begin{figure}[htbp]
    \scalebox{\figurescale}{\includegraphics[width=1\linewidth]{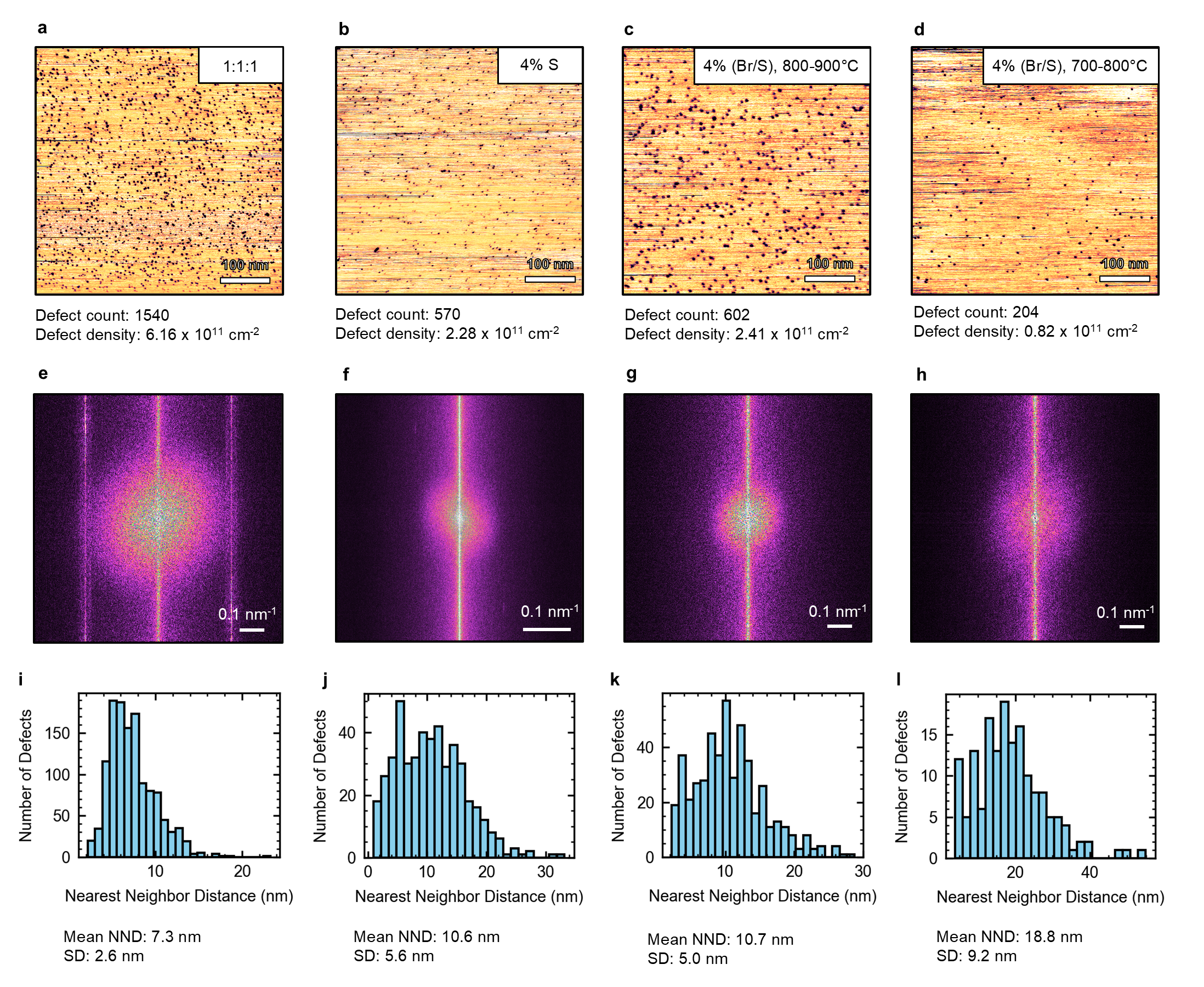}}
    \caption{\textbf{Spatial properties of one representative $500 \times 500$~nm$^2$ image from the stochiometric, 4\% S excess, 4\% Br/S excess at $800$-$900^\circ\mathrm{C}$, and 4\% Br/S excess at $700$-$800^\circ\mathrm{C}$ growth trial.} ~\textbf{a-d}, Representative images from stoichiomtetric growth, 4\% S growth, 4\% Br/S growth. and $700$-$800^\circ\mathrm{C}$ growth, respectively. ~\textbf{e-h}, 2D Fourier transforms showing isotropically distributed defects for each respective batch. ~\textbf{e} Contains additional vertical features due to noise in ~\textbf{a}. ~\textbf{i-l}, Nearest neighbor histograms, distance, and standard deviation for each respective batch.}
    \label{spatial_properties}
\end{figure}

The spatial distribution of defects can determine their interactions and are therefore of interest in our study. In Fig.~\ref{spatial_properties}, we demonstrate the spatial properties of one representative $500 \times 500$~nm$^2$ CrSBr image from the stoichiometric growth, 4\% S growth, 4\% Br/S growth at $800$-$900^\circ\mathrm{C}$, and 4\% Br/S growth at $700$-$800^\circ\mathrm{C}$ growth, respectively. These representative images are shown in  Fig.~\ref{spatial_properties}a-d. We focus on the 2D Fourier transform in Fig.~\ref{spatial_properties}e-h and nearest neighbor distance in Fig.~\ref{spatial_properties}i-l between defects as metrics. The nearest neighbor distance $d_{NN}$ was calculated in a range of $450 \times 450$~nm$^2$ in the center of each image to correct for any edge effects. From the Fourier transforms, we conclude that all batches show isotropically distributed defects. 

\subsection{C. Topographic Interaction with Electronic Defects}

\begin{figure}[htbp]
    \scalebox{\figurescale}{\includegraphics[width=1\linewidth]{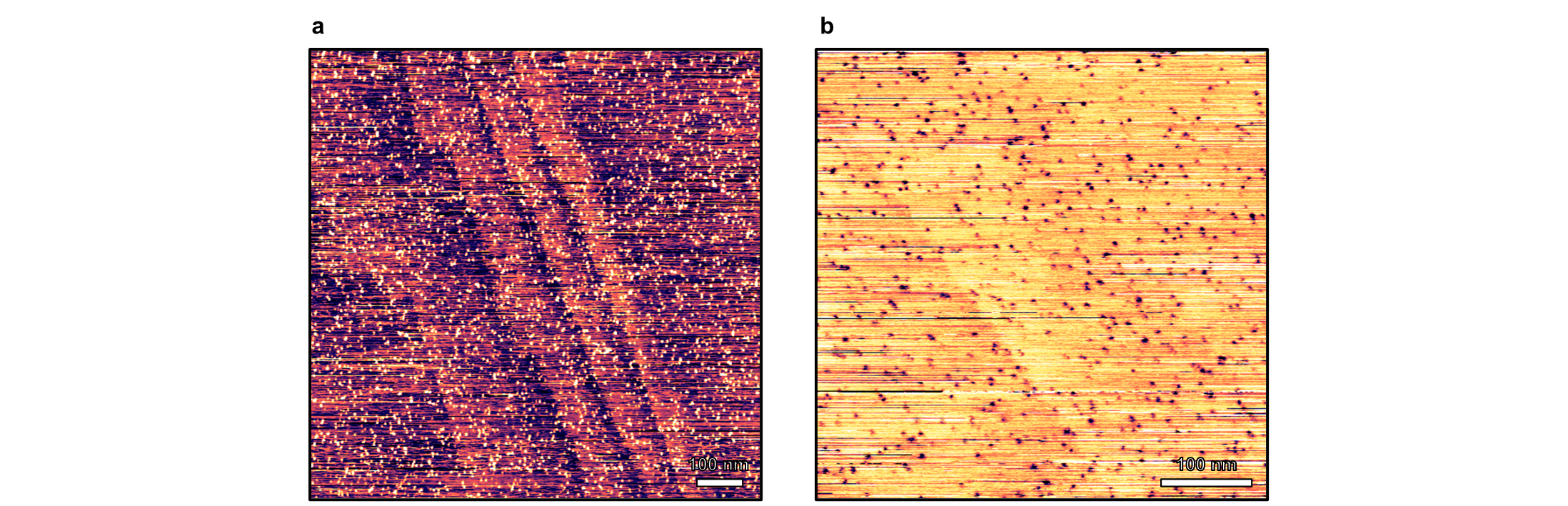}}
    \caption{\textbf{Defect interaction with topographic features.}
        \textbf{a}, $1 \times 1~\si{\micro\meter\squared}$ (4\% S excess growth) CrSBr image showing multiple atomic steps with defects not aligned around. \textbf{b}, $500 \times 500~\si{\nano\meter\squared}$ (4\% S excess growth) CrSBr showing an atomic step with defects not aligned around. 
	}
    \label{steps}
\end{figure}

In Fig.~\ref{steps} we present more images and the defect interactions with topographic features such as atomic steps. In Fig.~\ref{steps}a,b, our characteristic defect is not observed to cluster around atomic steps significantly, further supporting that they are not caused by topography (and/or mechanical strain from exfoliation). 

\begin{figure}[htbp]
    \scalebox{\figurescale}{\includegraphics[width=1\linewidth]{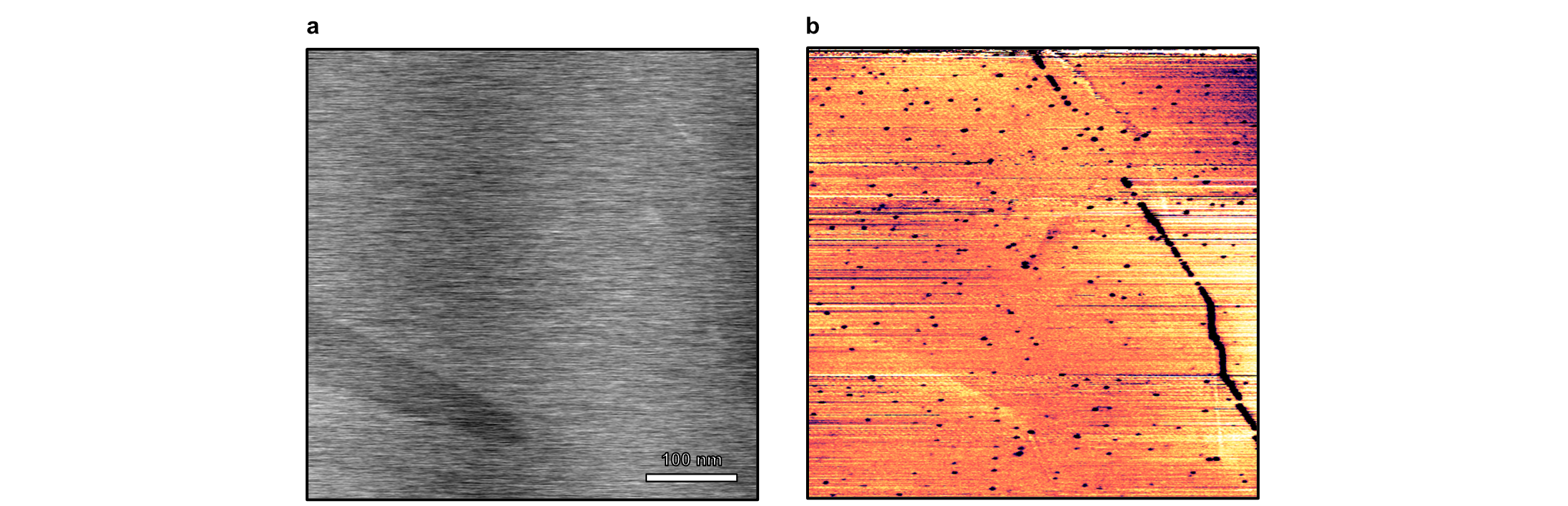}}
    \caption{\textbf{Ordering of defects without topographic correlation.}
        \textbf{a}, $500 \times 500$~nm$^2$ topographic AFM image from CAFM (contact) mode taken concurrently with the current image. \textbf{b}, Current image of CrSBr 4\% Br/S excess at ($700$-$800^\circ\mathrm{C}$ growth batch).
	}
    \label{ordering}
\end{figure}

In a smaller subset of images, we have observed ordering of defects in absence of correlated structure in the topographic image, and one $500 \times 500$~nm$^2$ example from the 4\% Br/S excess at $700$-$800^\circ\mathrm{C}$ growth batch is shown in Fig.~\ref{ordering}. Despite no correlated topography in the contact mode height image Fig.~\ref{ordering}a, the current image in Fig.~\ref{ordering}b shows defects arranged in a line pattern. We hypothesize that this example may be a dislocation that forms a linear arrangement of this electronic defect state.

\section{S3: Python Workflow for Defect Counting}

\begin{figure}[htbp]
    \scalebox{\figurescale}{\includegraphics[width=1\linewidth]{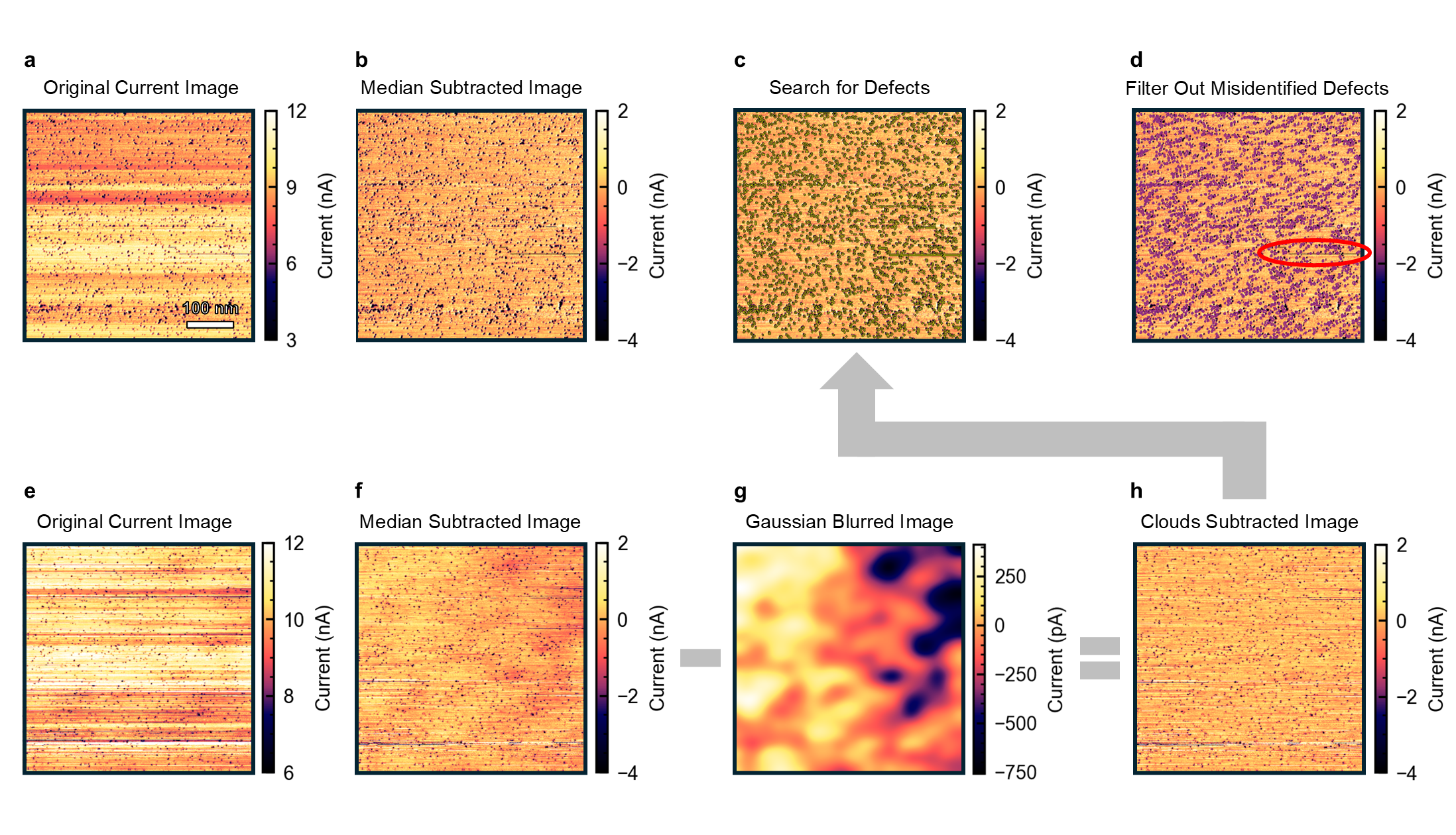}}
    \caption{{\textbf{CAFM defect counting workflows.}} 
        \textbf{a}, Raw data of defect image without strong background clouds. \textbf{b}, Median-subtracted image with clear defect appearance, no strong background clouds are observed. \textbf{c}, Defect mask applied to image, designed using a contour-finding algorithm using the OpenCV module. \textbf{d}, Defect mask filtered by defect size and aspect ratio to remove artifacts and misidentified defects. \textbf{e}, Raw data of defect image with strong background clouds. \textbf{f}, Median-subtracted image with strong background clouds observed. \textbf{g}, Gaussian blurred image. \textbf{h}, Image with background clouds effectively removed by subtracting gaussian blurred image in \textbf{g} from \textbf{f}. After this processing, the image goes through the same process as \textbf{c},\textbf{d}.  
	}
    \label{fig:counting}
\end{figure}

Fig.~\ref{fig:counting} depicts the analysis process that was used for visualizing and counting defects in CAFM images. 

A subset of $500 \times 500~\si{\nano\meter\squared}$ CAFM images displayed tens-of-nanometer-sized cloud-like features, clustered around areas of higher defect concentration. Two examples are shown in Fig.~\ref{topography_fig}a,b and Fig.~\ref{topography_fig}d,e. As a contrasting example, the CAFM images presented in the main manuscript do not show these strong background clouds. Further investigation is needed to determine the origin of these background clouds, but we hypothesize variations in local charging as origin.

For our image analysis, the choice of process depends on whether the image shows this strong cloud signal or not, since the presence of the clouds makes it difficult to detect and count the defects. 

If these features are absent, the process follows that outlined in Fig.~\ref{fig:counting}a-d. As is demonstrated, the raw image in Fig.~\ref{fig:counting}a undergoes a line-by-line median corrrection (subtraction) which removes striping artifacts, moving all lines to be at the same height. After this the defect appearance in Fig.~\ref{fig:counting}b is much clearer and defects can be easily identified. Once this is complete, the image is thresholded in grayscale and an OpenCV module in Python is used to detect defects as is shown in Fig.~\ref{fig:counting}c. A line-by-line normalization or global normalization of current values is applied to assist with thresholding and making the defects as identifiable and uniform as possible, used only for the purpose of counting the defect accurately. Then, as shown in Fig.~\ref{fig:counting}d, defects are filtered out on the basis of size and aspect ratio to eliminate misidentified defects that may arise from residual scan artifacts or noise in the data. From this information the defects were used to determine defect density, electronic information, and information in spatial distribution. 

If the image shows these strong clouds, the process first goes through Fig.~\ref{fig:counting}e-h and then through Fig.~\ref{fig:counting}c,d. In Fig.~\ref{fig:counting}f, striping artifacts are subtracted from the raw image in Fig.~\ref{fig:counting}e the same way they are in Fig.~\ref{fig:counting}b from Fig.~\ref{fig:counting}a. The cloud-like features can be visualized clearly at this point. In Fig.~\ref{fig:counting}g, a Gaussian blur, typically with $\sigma$ = (14.6 nm), is taken from Fig.~\ref{fig:counting}f using SciPy. The image in Fig.~\ref{fig:counting}g is then subtracted from Fig.~\ref{fig:counting}f to remove the effect of the clouds. The result of this subtraction is the image in Fig.~\ref{fig:counting}h, which clearly allows for more facile defect counting than Fig.~\ref{fig:counting}e or Fig.~\ref{fig:counting}f. Fig.~\ref{fig:counting}h then goes back through the process described in Fig.~\ref{fig:counting}c,d.

\section{S4: Electronic Role of Defect}

\subsection{A. Bias Series Analysis}

\begin{figure}[htbp]
    \scalebox{\figurescale}{\includegraphics[width=1\linewidth]{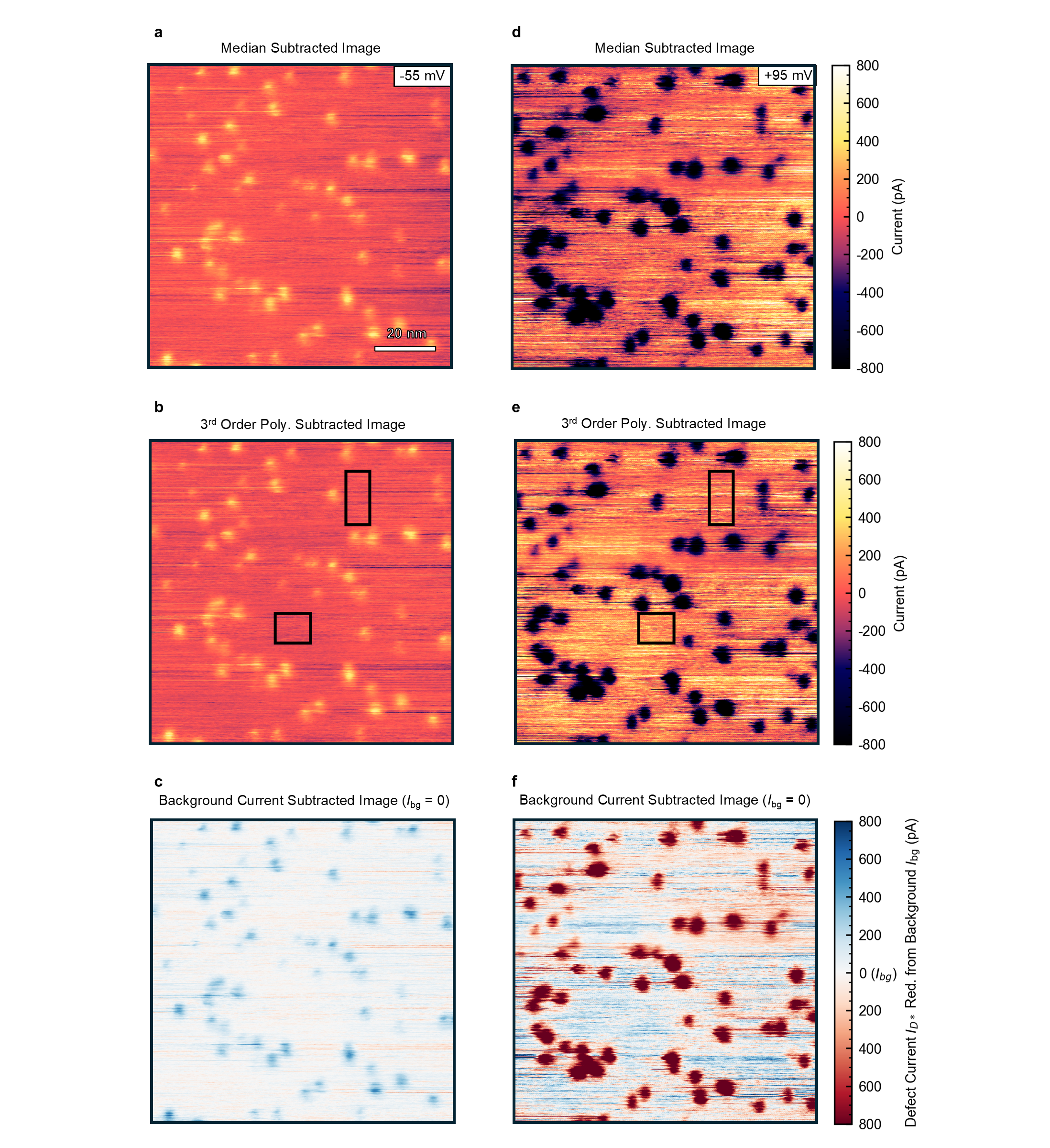}}
    \caption{\textbf{Absolute defect current magnitude reduction as a function of bias analysis for a small image.} \textbf{a}, $100 \times 100$~nm$^2$ -55 mV CAFM image (stoichiometric growth batch at $800$-$900^\circ\mathrm{C}$) that has undergone median subtraction. \textbf{b}, 3$^{rd}$ order polynomial background has been subtracted, the image visually has less tilt than in \textbf{a}. \textbf{c}, The CAFM image is represented with a white background showing the defect current magnitude reduction; this result is in Fig. 2a. \textbf{d}-\textbf{f}, Repeat these steps for a +95 mV image.
	}
    \label{bias_series_fig}
\end{figure}

We are interested in quantitative measures of the defect current with respect to the background (defect-free) current. We will now explain the analysis process for the images shown in Fig. 2a in the main manuscript that demonstrate the absolute current response of defects $I_{D*}$ in comparison to the background $I_{\mathrm{bg}}$. We begin with a $100 \times 100$~nm$^2$ -55 mV median-subtracted image (according to the process described in Fig.~\ref{fig:counting}) shown in Fig.~\ref{bias_series_fig}a. Then, we remove a polynomial background of order 3 to remove variations in the background, i.e. the defect-free areas, from scanner tilt. The results of this subtraction are shown in Fig.~\ref{bias_series_fig}b. Then, we identify two representative regions within the image that show no defects and find the mean of the values in those regions, which reflects the background current. We subtract this mean from the entirety of the image to set the background value $I_{\mathrm{bg}}$ to 0, since we aim to isolate the effect of the defects with respect to the background. This subtraction allows us to understand the absolute change in defect current $I_{D*}$ with respect to the background. We choose a color scale that highlights the differences in bias, from blue to red with white in the middle as no change with respect to the background. The final results are shown in Fig.~\ref{bias_series_fig}c. The process is again shown for a +95 mV image in Fig.~\ref{bias_series_fig}d-f. 

\subsection{B. Defect Current Reduction Maps}

\begin{figure}[htbp]
    \scalebox{\figurescale}{\includegraphics[width=1\linewidth]{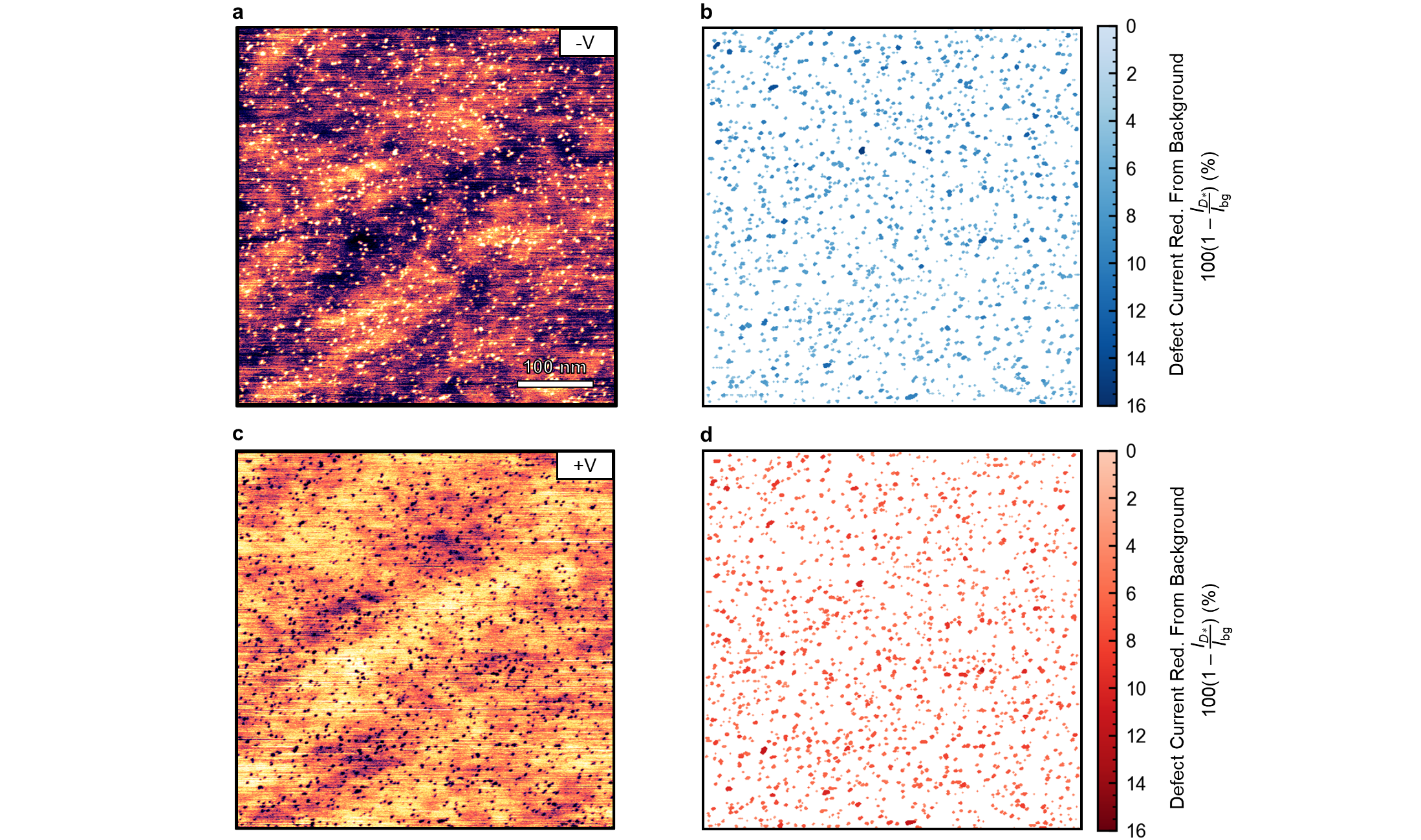}}
    \caption{\textbf{Flipped bias character and percent defect current magnitude reduction maps for a large image.}
        \textbf{a}, $500 \times 500$~nm$^2$ -V CAFM image (stoichiometric growth batch at $800$-$900^\circ\mathrm{C}$). 
        \textbf{b}, Corresponding masked defect map showing the current reduction of each defect relative to the background current of CrSBr.  
        \textbf{c}, +V CAFM image of the same area as in panel~\textbf{a} (stoichiometric growth batch at $800$-$900^\circ\mathrm{C}$). 
        \textbf{d}, Corresponding masked defect map showing the current reduction of each defect relative to the background current of CrSBr.  
	}
    \label{electronic_statistics_fig}
\end{figure}

Fig. 2b in the main manuscript shows the defect radius as a function of current magnitude reduction, $100\left(1 - \frac{I_{D*}}{I_{\mathrm{bg}}}\right)$, for two large-scale $500 \times 500$~nm$^2$ areas for the stoichiometric growth batch at $800$-$900^\circ\mathrm{C}$. Fig.~\ref{electronic_statistics_fig} shows shows the Fig.~\ref{electronic_statistics_fig}a,b negative and positive bias images and Fig.~\ref{electronic_statistics_fig}c,d map-like analysis that corresponds to the data shown in Fig. 2b. Visually, the defect size can be seen to correlate with the reduction in current magnitude from the background. The analysis workflow for this image processing is shown in Fig.~\ref{defect_reduction_fig}.

\begin{figure}[htbp]
    \scalebox{\figurescale}{\includegraphics[width=1\linewidth]{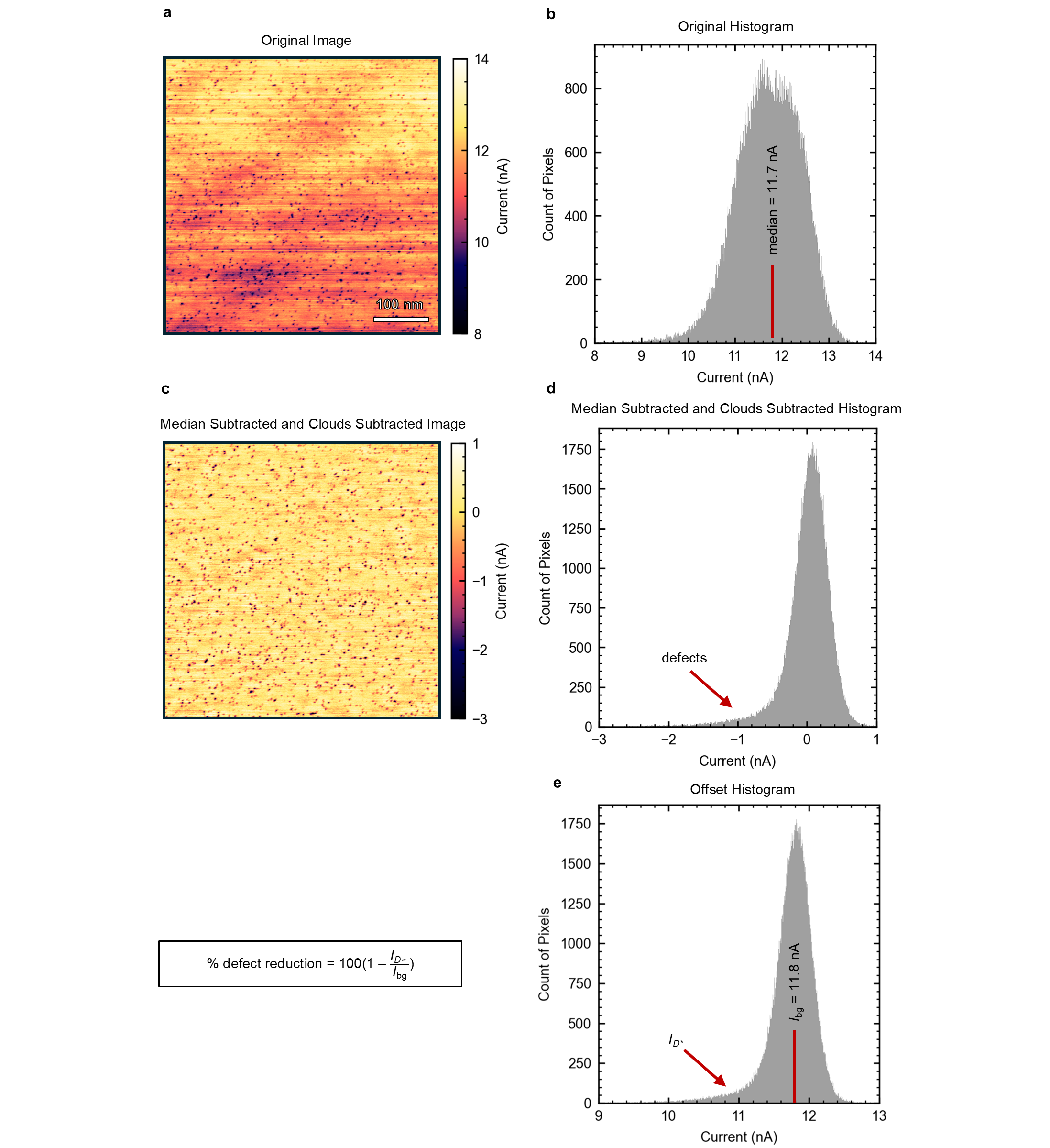}}
    \caption{\textbf{Percent defect current magnitude reduction from background calculation analysis for a large image.}
    \textbf{a}, Raw $500 \times 500$~nm$^2$ CAFM image (stoichiometric growth batch at $800$-$900^\circ\mathrm{C}$). \textbf{b}, Histogram of pixel frequency vs current values. \textbf{c}, Processed image and \textbf{d}, corresponding histogram. \textbf{e}, Shifted histogram from which defect reduction is calculated.
	}
    \label{defect_reduction_fig}
\end{figure}

\subsection{C. Defect Current Reduction Analysis}

We will now explain our quantitative analysis that produced Figs. 2b and Fig.~\ref{electronic_statistics_fig} using image processing as well as accompanying histograms. 

First, the histogram is generated for the raw image shown in Fig.~\ref{defect_reduction_fig}a and the median value of current is identified in Fig.~\ref{defect_reduction_fig}b. Then, a subtraction of scan lines and strong background clouds is performed according to the method described in Fig.~\ref{fig:counting}, and the result is shown visually in Fig.~\ref{defect_reduction_fig}c and in the histogram in Fig.~\ref{defect_reduction_fig}d. For our analysis, we subtract the presence of any strong background clouds (discussed in Fig.~\ref{fig:counting}) to keep consistency between images with and without background clouds. The histogram shows a tail on the left which represents the defect current values. We want to calculate the ratio of defects to background value $\frac{I_{D*}}{I_{\mathrm{bg}}}$, but any reasonable metric for the background, such as the median of the current values in Fig.~\ref{defect_reduction_fig}d, lies very close to 0 nA and thus does not produce a meaningful value.

To approximate $\frac{I_{D*}}{I_{\mathrm{bg}}}$, we shift the histogram in Fig.~\ref{defect_reduction_fig}d by the median value from Fig.~\ref{defect_reduction_fig}b to obtain the histogram in Fig.~\ref{defect_reduction_fig}e. We call the median value of the histogram in Fig.~\ref{defect_reduction_fig}e the background value $I_{\mathrm{bg}}$. The shifted defect values are what we now call $I_{D*}$. The defect current to background ratio is then calculated as $\frac{I_{D*}}{I_{\mathrm{bg}}}$, and the defect reduction in percent is $100\left(1 - \frac{I_{D*}}{I_{\mathrm{bg}}}\right)$. The process is shown here for the positive bias image in Fig.~\ref{electronic_statistics_fig}c but is identical for the negative bias image in Fig.~\ref{electronic_statistics_fig}a. The defect contours are identified and filtered using OpenCV as described in Fig.~\ref{fig:counting}c,d and plotted with their reduction values on a white background.

\subsection{D. Defects Under Repeated Scanning}

In Fig. 2a in the main manuscript, we demonstrated an instance where defects appear, disappear, and generally change appearance during repeated scanning. In Fig.~\ref{evolving_defects_fig_2}, we present an extension of this series. The image numbers correspond to the order in which the images were collected. Fig.~\ref{evolving_defects_fig_2}a is the last +95 mV image shown in Fig. 2a and Fig.~\ref{evolving_defects_fig_2}b-g follow sequentially. Out of approximately 70 defects in Fig. 2a (counted from the first image due to drift), only two changed their appearance over the first eight times scanning the area. Therefore, from this example we conservatively estimate the amount of defects altered from any image to be <5\% as long as the area is not scanned excessively. For our analysis, we typically take the first or second image to scan over an area, depending on image quality. 

\begin{figure}[htbp]
    \scalebox{\figurescale}{\includegraphics[width=1\linewidth]{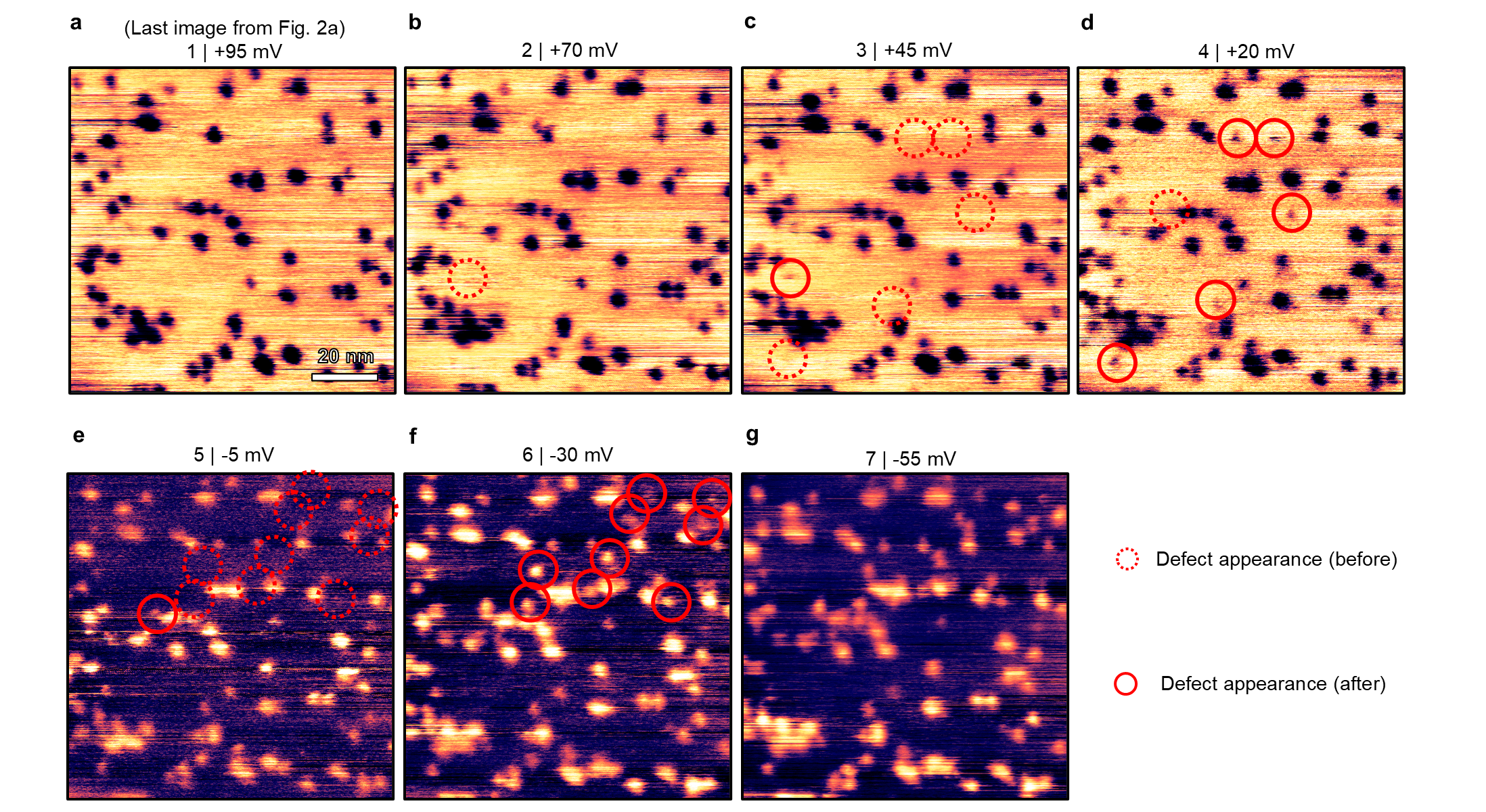}}
    \caption{\textbf{Continuation of scanning over area shown in Fig. 2a highlighting defect evolution.} \textbf{a}, Is a repeat of the last image in Fig. 2a taken at +95 mV for clarity. The images are further presented and numbered in the order they were taken: \textbf{b}, at +70 mV, \textbf{c}, at +45 mV, \textbf{d}, at +20 mV, \textbf{e}, at -5 mV, \textbf{f}, at -30 mV, \textbf{g}, at -55 mV. The scalebar shown in \textbf{a} applies to all images.
	}
    \label{evolving_defects_fig_2}
\end{figure}

\begin{figure}[htbp]
    \scalebox{\figurescale}{\includegraphics[width=1\linewidth]{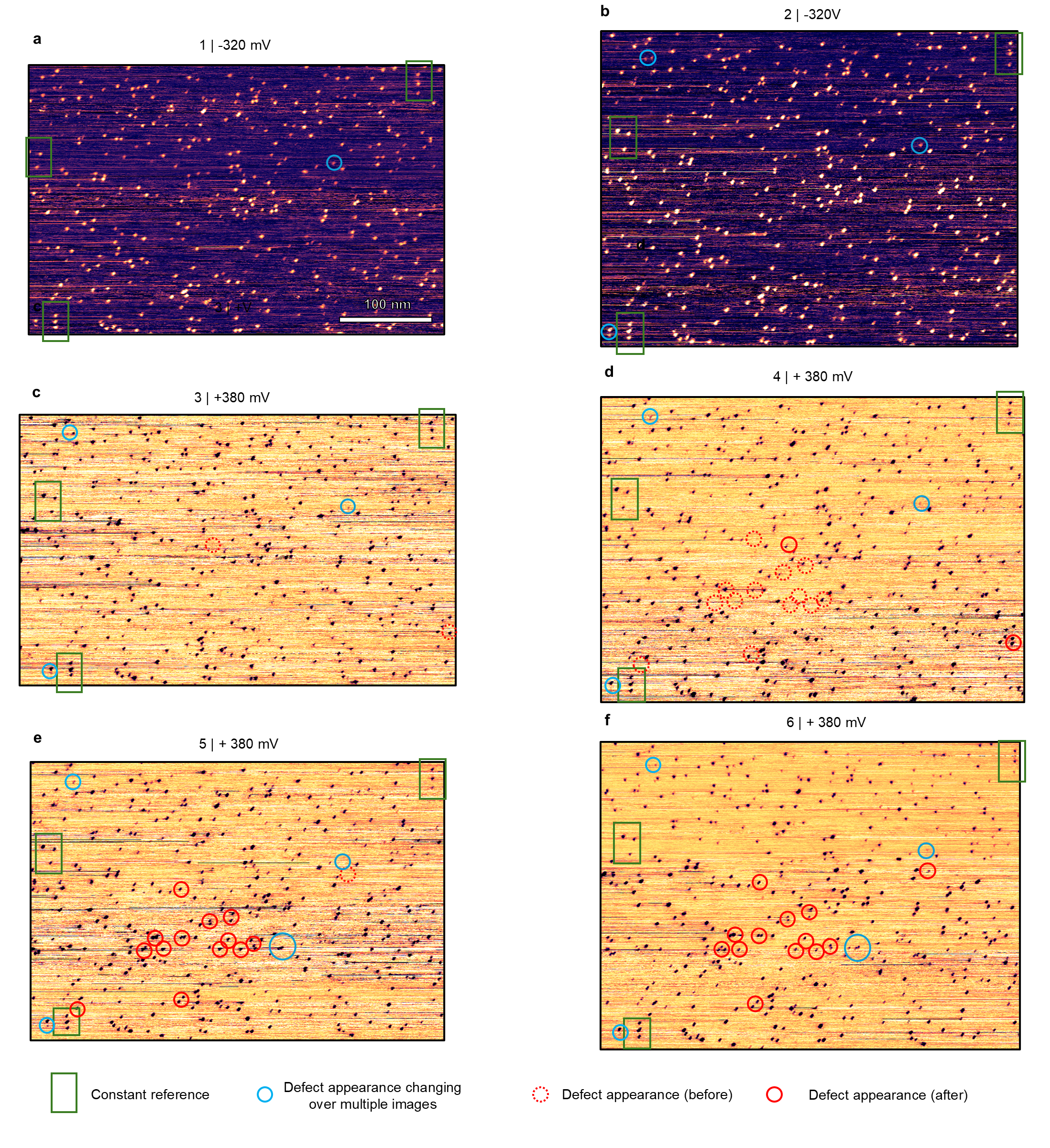}}
    \caption{\textbf{Six sequential images of an area highlighting defect evolution during scanning.} \textbf{a,}\textbf{b}, Constant -320 mV images taken before \textbf{c-f}, constant +380 mV images. The scalebar shown in \textbf{a} applies to all images; areas appear different sizes due to drift and scan distortion.
	}
    \label{evolving_defects_fig}
\end{figure}

In Fig.~\ref{evolving_defects_fig}, we present another distinct example of six images (from the 4\% S excess batch) taken sequentially for the purpose of examining appearing/disappearing defects. The first two Fig.~\ref{evolving_defects_fig}a,b are collected under constant -320 mV and the last four Fig.~\ref{evolving_defects_fig}c-f are collected under constant +380 mV. Due to sample drift and scan distortions, a consistent area is selected and compared between all images.

Three defect clusters are marked with a green square as a guide to the eye between images. Across the images, select areas where defects change (such as appearing, disappearing, and/or seeming to split/merge) over a series of images are marked with a blue circle. Separately, areas where defects specifically appear from one image to another are marked, as defect appearance is more commonly observed than defect disappearance. The image(s) after the defect appears is marked with a solid red circle and the image(s) before it appears is marked with a dotted red circle. 

Throughout both image series shown in Fig.~\ref{evolving_defects_fig_2} and Fig.~\ref{evolving_defects_fig}, we find that more defects appear the longer scanning occurs (the more frequent the same area is scanned). Consistent with this observation, we hypothesize that the defect appearance can originate from charge effects. The tip locally injects electrons or draws electrons from the vicinity of the defect, providing a reservoir of charges that can alter the charge state of a defect. Changes in charge state can drastically change the electronic property of the defect, metallic or insulating, and therefore the contrast mechanism in CAFM.

\section{S5: Defect Homogeneity Across and Between Flakes}

We are now interested in examining the variation in density between different areas within one exfoliated flake as well as between areas on different exfoliated flakes, as we observed evidence of inhomogeneity during growth for all elemental excess and varied temperature CrSBr batches. This investigation provides further information about the homogeneity of the growth.

\begin{figure}
    \scalebox{\figurescale}{\includegraphics[width=1\linewidth]{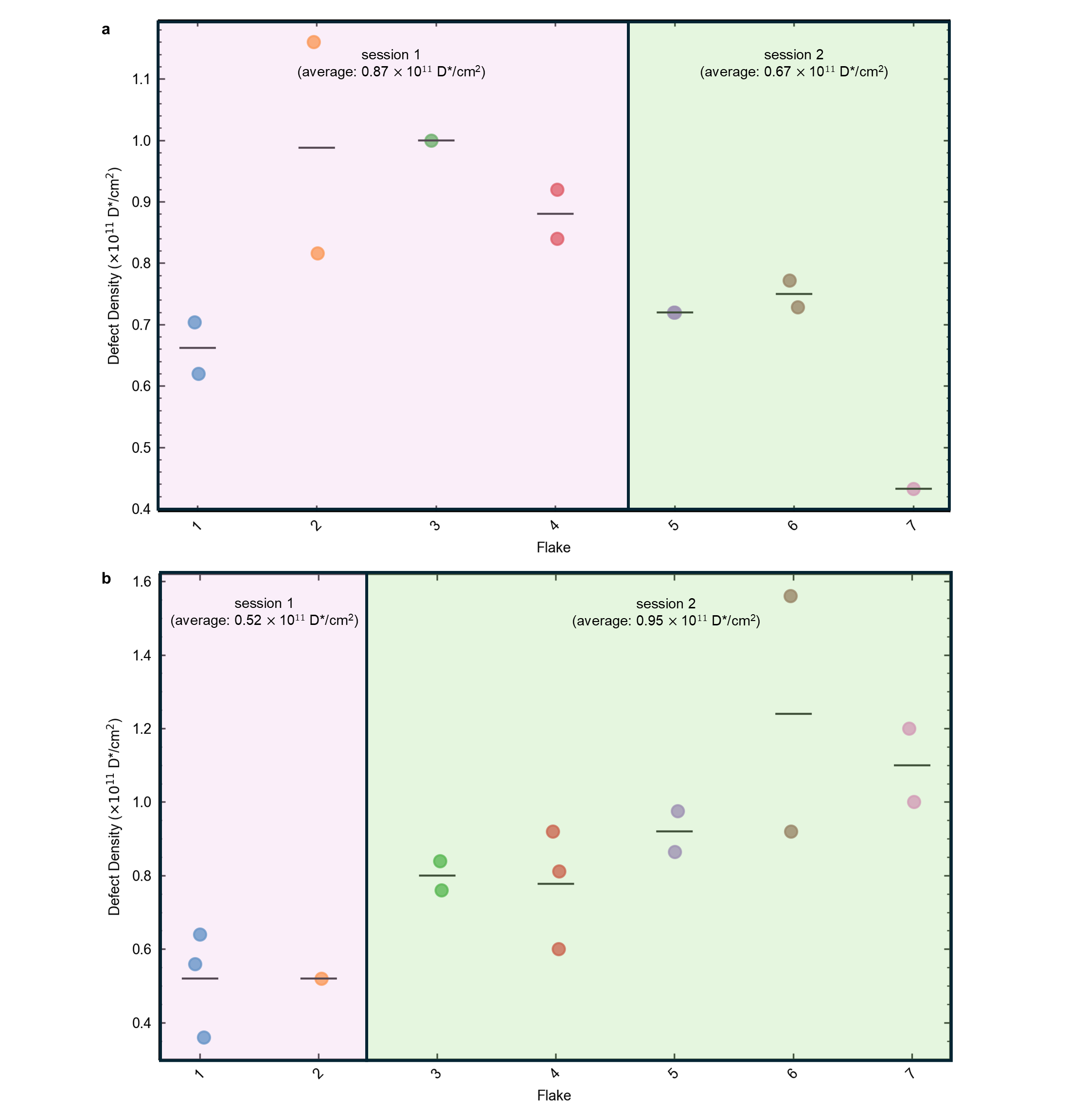}}
    \caption{\textbf{Flake-by-flake comparison of measured densities from the Br/S 4\% excess at $700$-$800^\circ\mathrm{C}$ and $750$-$850^\circ\mathrm{C}$ growth batches.}  \textbf{a}, Densities from seven flakes from the $700$-$800^\circ\mathrm{C}$ temperature gradient trial taken over two sessions, highlighted in purple and green. \textbf{b}, Densities from seven flakes from the $750$-$850^\circ\mathrm{C}$ temperature gradient trial taken over two sessions, highlighted in purple and green. Each data point represents one $500 \times 500$~nm$^2$ image, and black lines represent flake averages. Each session used a new 2D-like sample.
	}
    \label{within_between_fig}
\end{figure}

To this end, we take the results from the varied temperature growth as a case study (both the Br/S 4\% excess at $700$-$800^\circ\mathrm{C}$ growth batch and the Br/S 4\% excess at $750$-$850^\circ\mathrm{C}$ growth batch). In Fig.~\ref{within_between_fig}, we present graphs showing the density from each $500 \times 500$~nm$^2$ image within each exfoliated flake of the Fig.~\ref{within_between_fig}a $700$-$800^\circ\mathrm{C}$ growth and the Fig.~\ref{within_between_fig}b $750$-$850^\circ\mathrm{C}$ growth as a visual guide for how the density varies across all data points collected.

To provide a quantitative investigation, we first calculate the difference in the density between each individual image. We calculate this quantity $\Delta \sigma$ within each flake (for flakes that have multiple images), and average the values to get the metric $\langle \Delta \sigma \rangle_{\mathrm{within}}$. Then, we calculate the average differences between images on different flakes to get the $\langle \Delta \sigma \rangle_{\mathrm{between}}$. We calculate the $\langle \Delta \sigma \rangle_{\mathrm{ratio}}$ as the ratio of the $\langle \Delta \sigma \rangle_{\mathrm{within}}$ to the $\langle \Delta \sigma \rangle_{\mathrm{between}}$. The result of this metric gives a quantitative measure of how much closer the values within flakes are as compared to the values between flakes.

For the $700$-$800^\circ\mathrm{C}$ growth, we calculate the $\langle \Delta \sigma \rangle_{\mathrm{within}}$ metric to be $ 1.1 \times10^{10}\,\mathrm{cm}^{-2}$, the $\langle \Delta \sigma \rangle_{\mathrm{between}}$ to be $2.2 \times10^{10}\,\mathrm{cm}^{-2}$, and the $\langle \Delta \sigma \rangle_{\mathrm{ratio}}$ to be 0.5. From the $750$-$850^\circ\mathrm{C}$ growth, the $\langle \Delta \sigma \rangle_{\mathrm{within}}$ metric was calculated to be $2.2 \times10^{10}\,\mathrm{cm}^{-2}$, the $\langle \Delta \sigma \rangle_{\mathrm{between}}$ metric was calculated to be $3.4 \times10^{10}\,\mathrm{cm}^{-2}$, and the $\langle \Delta \sigma \rangle_{\mathrm{ratio}}$ metric was calculated to be 0.65. 

For both the $700$-$800^\circ\mathrm{C}$ batch and the $750$-$850^\circ\mathrm{C}$ batch, the $\langle \Delta \sigma \rangle_{\mathrm{ratio}}$ metric is less than one. We interpret this to mean that images taken within the same flake have a more similar defect density. This finding is expected, considering that the defect density is tied to the growth temperature and there is a temperature gradient within each flake as they grow along the side of the ampoule. One conclusion from these results is that measuring multiple different areas is necessary, since the defect density can vary between them. 

\begin{figure}
    \scalebox{\figurescale}{\includegraphics[width=1\linewidth]{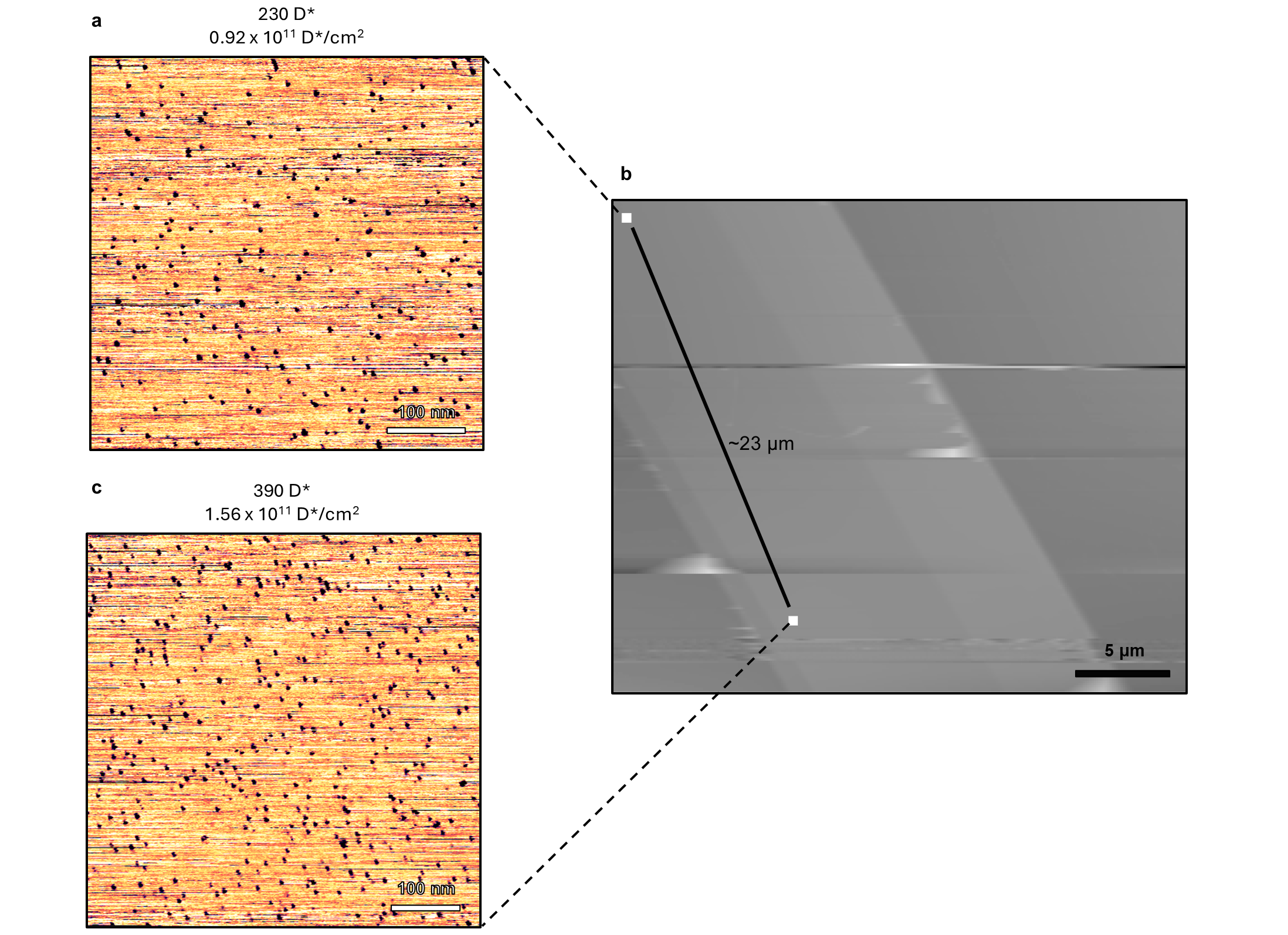}}
    \caption{\textbf{Two $500 \times 500$~nm$^2$ images collected $\sim$ 23 $\si{\micro\meter}$ away from each other on the same exfoliated flake of the Br/S 4\% excess at $750$-$850^\circ\mathrm{C}$ growth batch.} The two CAFM images are shown in \textbf{a} and \textbf{b}. The tapping mode topographic image from which they are referenced is in \textbf{c}.
	}
    \label{within_between_ex1}
\end{figure}

In Fig.~\ref{within_between_ex1} we present one visual example of the variation among areas on one flake. Fig.~\ref{within_between_ex1}a,b are the two $500 \times 500$~nm$^2$ images from flake 6 of the Br/S 4\% excess at $750$-$850^\circ\mathrm{C}$ growth as shown, with densities of $0.92\times 10^{11}\,\mathrm{cm}^{-2}$ and $1.56 \times10^{11}\,\mathrm{cm}^{-2}$ respectively. These areas are approximately 23 $\SI{}{\micro\meter}$ away from each other and there is a 50\% difference in concentration. Figure~\ref{within_between_ex1}c shows the topographic image of the flake from which they were taken on.

\clearpage
\section{S6: Lattice Distortion due to Vacancy Defect with DFT}
\begin{figure}[h]
    \centering
    \scalebox{\figurescale}{\includegraphics[width=1\linewidth]{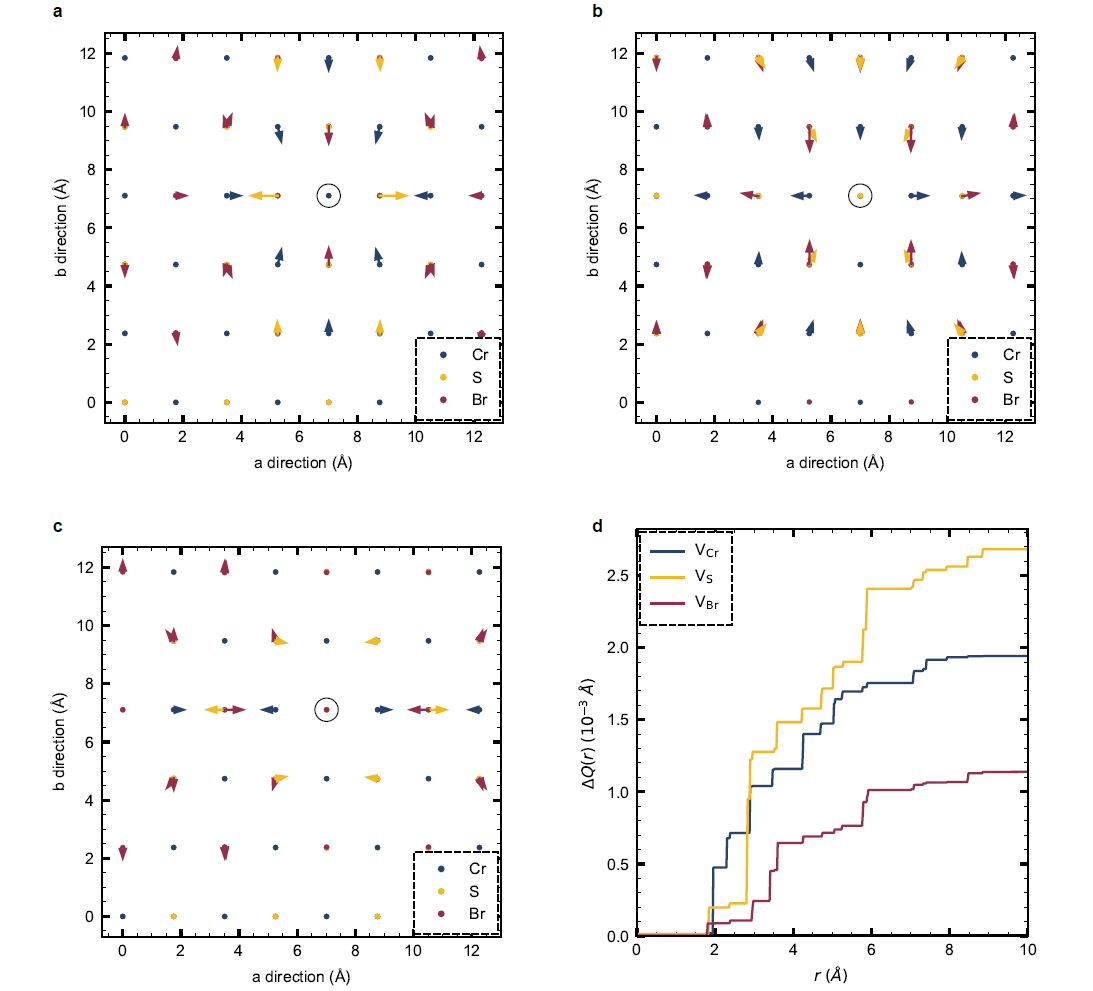}}
    \caption{\textbf{Lattice distortion due to vacancies in CrSBr supercell.} Top view of unit cell of with \textbf{a}, Cr, \textbf{b}, S, \textbf{c}, Br vacancy in CrSBr supercell with arrows (magnified by factor 3) showing displacement from equilibrium structure. \textbf{d}, Summed displacement $\Delta Q$ as a function of radius $r$ around the vacancy center.}
    \label{SI_lattice}
\end{figure}

Similarly to the charge redistribution induced by different point defects as shown in Fig 6a-c in the main manuscript, here we look at the local lattice distortions.
Fig. ~\ref{SI_lattice}a-c displays the resulting atomic displacements for the Cr, S and Br vacancy structures, respectively.
The displacements are symmetric with respect to the x- and y-mirror planes through the defect site. Notably, the Br vacancy shows anisotropic behavior as the displacement of the atoms is largest in the $a$ direction.
The displacement of the atoms is quantified by evaluating $\Delta Q (r) = \sum_{r'_i<r}|\tau_i(r'_i)|$ with $\tau_i(r'_i)$ the displacement of atom $i$ at position $r'_i$ as shown in Fig. ~\ref{SI_lattice}d.
The total summed displacement of the atoms for the S vacancy structure is largest, as is also the case for the $\Delta\rho$.
The $\Delta Q$ has a large jump around $r=2~\text{\AA}$ for the Cr vacancy, whereas for the S and Br vacancy, this jump happens at larger $r$.
This indicates a more local effect of the Cr vacancy compared to S and Br, in line with the results from the charge redistribution. \\

\FloatBarrier
\centering

\textbf{References}
\bibliographystyle{apsrev}
\bibliography{full}